\documentclass[
reprint,
jcp,
aip,
]{revtex4-1}

\usepackage{amssymb}
\usepackage{placeins}
\usepackage{bm}
\usepackage{amsmath}
\usepackage{graphicx}
\usepackage{mathtools}
\usepackage{gensymb}
\usepackage[table]{xcolor}
\usepackage{array}
\usepackage{amsfonts}
\usepackage{multirow}
\usepackage[utf8]{inputenc}
\usepackage[T1]{fontenc}
\usepackage{lmodern}
\usepackage[outdir=./,update]{epstopdf}
\usepackage[english]{babel}
\usepackage{bm}

\begin{document}


\title{\textit{Ab initio} study of reactive collisions between Rb($^{2}S$) or Rb($^{2}P$) and OH$^{-}$($^{1}\Sigma^{+}$)}

\author{Milaim Kas}
\email{milakas@ulb.ac.be}
\author{Jérôme Loreau}
\author{Jacques Liévin}
\author{Nathalie Vaeck}%
\affiliation{
 Service de Chimie Quantique et Photophysique (CQP)\\
 Université libre de Bruxelles (ULB), Brussels, Belgium 
}

\date{\today}

\begin{abstract}

A theoretical rate constant for the associative detachment reaction Rb($^{2}S$)$+$OH$^{-}$($^{1}\Sigma^{+}$)$\rightarrow \,$ RbOH($^{1}\Sigma^{+}$)$+\,e^{-}$ of 4$\,\times$10$^{-10}$cm$^{3}$s$^{-1}$ at 300 K has been calculated. This result agrees with the experimental rate constant of 2$^{+2}_{-1}\,\times10^{-10}$cm$^{3}$s$^{-1}$ obtained by Deiglmayr \textit{et al.} (Phys. Rev. A 86, 2012) for a temperature between 200 K and 600 K. A Langevin-based dynamics which depends on the crossing point between the anion (RbOH$^{-}$) and neutral (RbOH) potential energy surfaces has been used. The calculation were performed using the ECP28MDF effective core potential to describe the rubidium atom at the CCSD(T) level of theory and extended basis sets. The effect of ECPs and basis set on the height of the crossing point, and hence the rate constant, has been investigated. The temperature dependence of the latter is also discussed. Preliminary work on the potential energy surface for the excited reaction channel Rb($^{2}P$)+OH$^{-}$($^{1}\Sigma^{+}$) calculated at the CASSF-icMRCI level of theory is shown. We qualitatively discuss the charge transfer and associative detachment reactions arising from this excited entrance channel.
 
\end{abstract}

\maketitle


\section{\label{sec:int}Introduction}

The last decade has been the subject of numerous advances in the study of the dynamics of cold atoms and molecules, leading to the opening of new avenues in various branches of physics and chemistry \cite{Carr2009,Dulieu2011,Krems2010}. The applications range from precision spectroscopy \cite{Loh2013} to quantum control of chemical reactions \cite{Quemener2012}. New research fields have been developed in order to control and cool atomic and molecular species, allowing the possibility to investigate their interactions at low temperature. Depending on the experimental method, the translational, rotational, vibrational or hyperfine energy can correspond to temperatures below 1 K. Neutral and ionic atoms can now be routinely trapped and laser-cooled to the micro Kelvin regime \cite{Krems2010}. A broad range of methods such as laser cooling, Stark deceleration, radio-frequency trap (rf trap), buffer gas or sympathetic cooling have been developed to cool and/or trap molecular species. Molecular ions can be trapped using an rf trap \cite{Wester2009a} which allows subsequent cooling by collision with cold gas or with laser cooled atoms. The former method, using standard cryostats, is limited to temperatures above 4 K \cite{Pearson1995} whereas in the latter, translational temperature down to a few millikelvins can be reached using laser cooled atomic ions \cite{Krems2010}. The use of neutral atoms, implying the collision of much closer encounters should result in the cooling of internal degrees of freedom, therefore reducing the population of high rotational and vibrational states of the molecular ion. Such a scheme, called an hybrid atom-ion trap, consists of an rf-trap superimposed with a magneto optical trap (MOT) \cite{Hudson2009}. In such an environment, elastic, inelastic and reactive collisions can take place, leading to a loss of ions if the kinetic energy release exceeds the depth of the trap or if the ions lose their charge. The co-trapping of Rb and OH$^{-}$ has been the subject of several theoretical studies \cite{Byrd2013,Tacconi2009a,Gonzalez-Sanchez2015,Gonzalez-Sanchez2008,Gonzalez-Sanchez2015a} and is currently under experimental investigation by the HAItrap group in the university of Heidelberg from which the first results have been published \cite{Deiglmayr2012}. This system is of particular interest since the dynamics of anions, specially at low temperature, can exhibit non-standard behaviour \cite{Otto2008}. Moreover, quantum chemistry calculations involving anions have also proven to be challenging \cite{Simons2008a}.\\
In the co-trapping experiment \cite{Deiglmayr2012} of Rb and OH$^{-}$, a loss of OH$^{-}$ has been observed and attributed to the associative detachment reaction Rb($^{2}$S)+OH$^{-}$($^{1}\Sigma^{+}$) $\rightarrow \,$ RbOH($^{1}\Sigma^{+}$) + $e^{-}$, and a translational temperature of 200-600 K has been estimated for OH$^{-}$. The first theoretical result was obtained by Byrd et al. \cite{Byrd2013} using a Langevin model based on the calculated crossing point between the anion and neutral potential energy surfaces (PES). They showed that the associative detachment reaction can only occur when higher excited vibrational states of OH$^{-}$ are taken into account. However, the presence of vibrational excited OH$^{-}$ in the trap is very unlikely. In fact, the vibrational frequency of OH$^{-}$ is 3556 cm$^{-1}$ ($0.44$ eV) \cite{Branscomb1966}, which corresponds to an equivalent temperature energy ($E$=$k_{b}T$) of 5116 K. At 300 K the population ratio of the first vibrational excited state ($v$=1) is about 4$\times10^{-8}$, which is negligible. One could argue that the OH$^{-}$ molecules are produced in a plasma discharge, which is known to produce vibrational hot species \cite{Simons2008a}. However, the vibriationally excited OH$^{-}$ should quickly (about ten ms \cite{Byrd2013}) decay to their vibrational ground state by spontaneous emission. Moreover, the OH$^{-}$ molecules are thermalized by collision with argon atoms at room temperature before being trapped \cite{Deiglmayr2012}. The trapped OH$^{-}$ molecules should therefore be in their vibrational ground state when colliding with the Rb atoms. \\ 
Since the optical cycle used in the MOT involves rubidium in its first excited electronic state $^2P$ ($^{2}P_{3/2}$ when taking into account the fine structure), reactions from the excited entrance channel Rb($^{2}P$)+OH$^{-}$($^{1}\Sigma^{+}$) are also interesting to study and could lead to some unusual effects \cite{Hall2012}. \\
The present paper is structured as follows: in section \ref{sec:Sym} we start by defining the different symmetry correlations for the RbOH and RbOH$^{-}$ systems. In section \ref{sec:AD} we recalculate a portion of the potential energy surface (PES) of the RbOH and RbOH$^{-}$ species using three different effective core potentials (ECP) for the rubidium atom. The crossing point between the anion and neutral curve, corresponding to the autodetachment region, has been obtained from the best PES. We recalculate the rate constant for the associative detachment reaction using the model suggested by Byrd et al. and the results of the newest ECP. In the last part of the paper (section \ref{sec:Exc}), we investigate the excited reaction channel.             

\section{\label{sec:Sym}Symmetry considerations}

\begin{figure}
\begin{center}
\includegraphics [width=8.5cm]{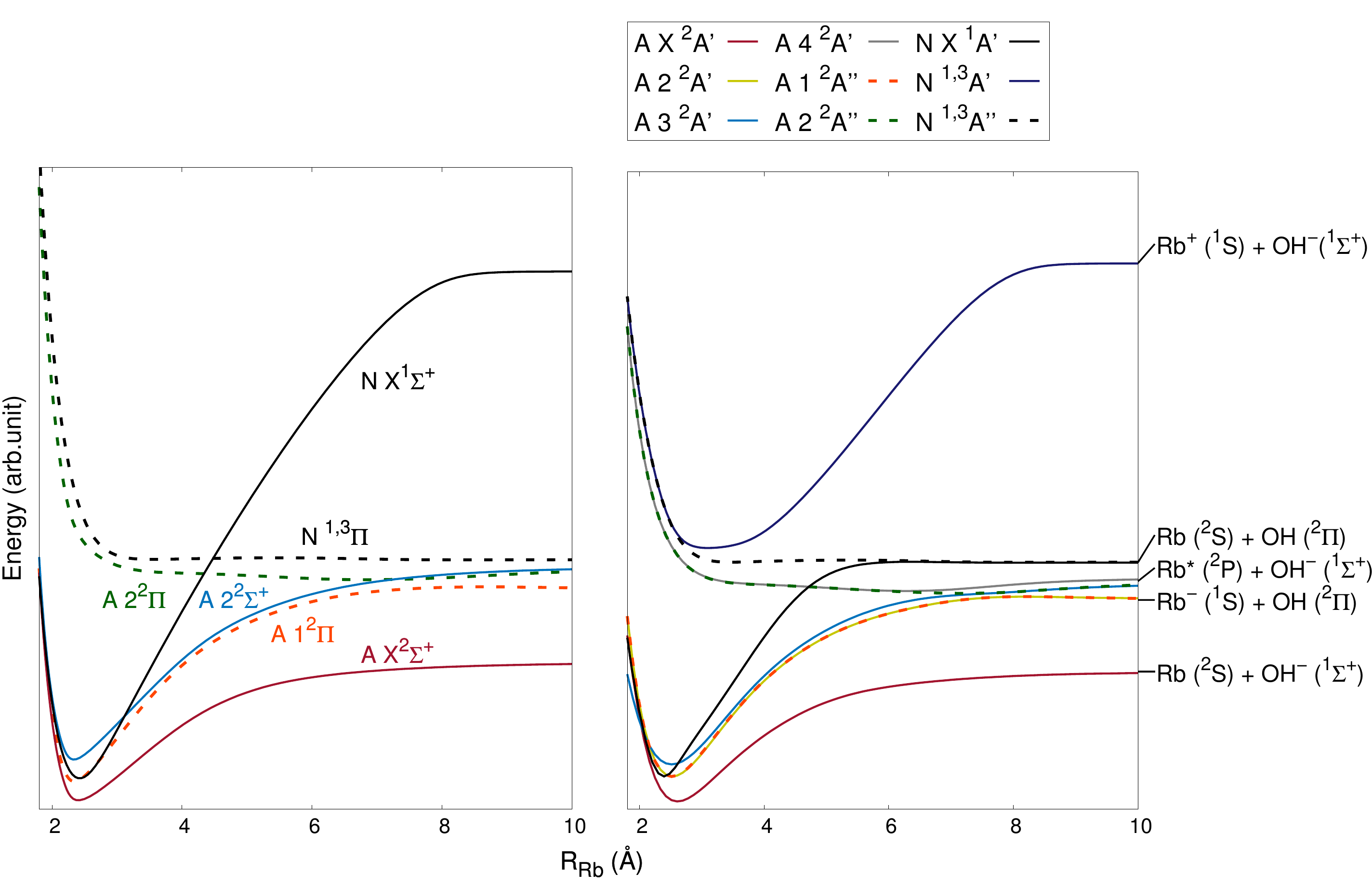}
\end{center}
\caption{\small Schematic potential energy curves for the low-lying states of the anionic RbOH$^{-}$ molecule and for the neutral RbOH state that correlate to the ionic and covalent dissociation limits. The left and right panels correspond to the linear and bent geometry respectively. The dissociative states are shown. Neutral and anionic states are identified by the labels N and A, respectively. The $\Pi$ and corresponding $A''$ states are indicated by dashed lines whereas the $\Sigma$ and corresponding $A'$ states are indicated by solid lines.}
\label{fig:PEC_tot}
\end{figure}

The ground state of RbOH$^{-}$ is defined by the $^{2}\Sigma^{+}$ term symbol in the linear $C_{\infty v}$ symmetry which becomes $^{2}A'$ in the nonlinear case ($C_{s}$ point group). This molecular state correlates to the dissociative channel Rb($^{2}S$)+OH$^{-}\,(^{1}\Sigma^{+})$, both products being in their ground states. The ground state of the neutral RbOH molecular species is defined by the $^{1}\Sigma^{+}$ term symbol in $C_{\infty v}$ and $^{1}A'$ in $C_{s}$ symmetry. This state is correlated to the covalent Rb($^{2}S$)+OH($^{2}\Pi$) dissociation channel for bent geometries and to the ionic Rb$^{+}$($^{1}S$)+OH$^{-}$($^{1}\Sigma^{+}$) dissociation channel at linear geometry \cite{Lara2007}.\\ 
The first electronic excited state of Rb is a $^{2}P$ state, which correlates with the $^{1}\Sigma^{+}$ ground state of OH$^{-}$ to form a $^{2}\Sigma^{+}$ and a $^{2}\Pi$ excited molecular states at linear geometry. At bent geometry these states split into two $^{2}A''$ and one $^{2}A'$ states. For the sake of completeness we also have to take into account the charge transfer channel Rb$^{-}$($^{1}S$)+OH($^{2}\Pi$) which lies around 0.24 eV below the Rb($^{2}P$)+OH$^{-}$($^{1}\Sigma^{+}$) channel (see section \ref{sec:Exc}) and lead to the $1\,^{2}\Pi$ molecular state (which becomes $2\,^{2}A'+\,1\,^{2}A''$ at bent geometries). These Wigner-Witmer correlations for the anion along with those for the neutral RbOH molecular specie are summarized in \tablename{~\ref{tabsym}}. The energies of the different channels are given relative to Rb($^{2}S$)+OH$^{-}$($^{1}\Sigma^{+}$). It should be pointed out that several other dissociation limits are located between the Rb($^{2}S$)+OH($^{2}\Pi$) and Rb$^{+}$($^{1}S$)+OH$^{-}$($^{1}\Sigma^{+}$) channels \cite{Byrd2013} but are not accessible in the energy range considered here. A schematic overview of the potential energy curves (PEC) at linear and bent geometry for the different states appearing in \tablename{~\ref{tabsym}} is shown in \figurename{~\ref{fig:PEC_tot}}. The curves have been obtained from separate MRCI calculations for the neutral and anionic system and shifted to match the experimental energies at dissociation and are therefore only shown for a better understanding of the different states of interest.       
\begin{table}[h]
\caption{\small Wigner-Witmer correlation rules for the Rb-OH neutral and anionic system. Experimental values for the energy of the different dissociation channels are obtained from the electroaffinity of OH \cite{Smith1997} and Rb \cite{Frey1978}, the electronic excitation energy and ionization energy of Rb \cite{Sansonetti2006} and are given relative to the Rb($^{2}S$)+OH$^{-}$($^{1}\Sigma^{+}$) channel.}
\resizebox{\columnwidth}{!}{
\def\arraystretch{1.3}
\begin{tabular}{llccc}
\hline
\hline
& & \multicolumn{2}{c}{Molecular states} &  \\
\cline{3-4}
& Dissociation limit & $C_{\infty v}$(linear) & $C_{s}$ (bent) & Energy (eV) \\
\hline
\multirow{2}{*}{Neutral} & Rb$^{+}$($^{1}S$)+OH$^{-}$($^{1}{\Sigma^{+}}$) & $X\,^{1}\Sigma^{+}$ & 2$\,^{1}A'$ & 4.18 \\
& Rb($^{2}S$)+OH($^{2}\Pi$) & 1$\,^{1,3}\Pi$ & $X\,^{1}A' \oplus$ 1$\,^{1,3}A''\oplus$ 1$\,^{3}A'$ & 1.83 \\
\hline
\multirow{3}{*}{Anion} & Rb$^{*}$($^{2}P$)+OH$^{-}$($^{1}{\Sigma^{+}}$) & 2$\,^{2}\Sigma ^{+} \oplus$ 2$\,^{2}\Pi   $  & 3$\,^{2}A' \oplus$ 4$\,^{2}A'$ $\oplus$ 2$\,^{2}A''$ & 1.56-1.59$^{1}$  \\
& Rb$^{-}$($^{1}S$)+OH($^{2}{\Pi}$) & 1$\,^{2}\Pi$ & 2$\,^{2}A' \oplus$ 1$\,^{2}A''$ & 1.33 \\
& Rb($^{2}S$)+OH$^{-}$($^{1}{\Sigma^{+}}$) & $X\,^{2}\Sigma^{+}$ & $X\,^{2}A'$ & 0 \\
\hline
\hline
\multicolumn{5}{l}{\footnotesize $^{1}$ Excitation energy relative to the $^{2}P_{1/2}$ and $^{2}P_{3/2}$ fine states of Rb, respectively.}
\end{tabular}
}
\label{tabsym}
\end{table}

\section{\label{sec:AD}Collision with ground state rubidium: associatve detachment}
\subsection{\label{sec:AD_method}Computational method}
\begin{figure}[h]
\begin{center}
\includegraphics[width=4cm]{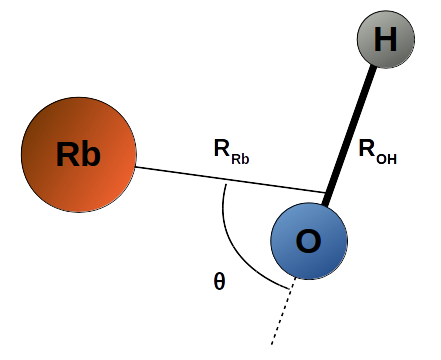}
\end{center}
\caption{\small Jacobi coordinates for the Rb-OH molecular system.}
\label{fig:RbOH}
\end{figure}

All calculations were performed on the Hydra and Vega clusters of the ULB/VUB using the \begin{small}MOLPRO\end{small} 2012 package \cite{MOLPRO}. Jacobi coordinates have been used to describe the geometry of the Rb-OH system. $R_\text{{Rb}}$ is the distance between the Rb atom and the center of mass of the OH molecule, $R_\text{{OH}}$ is the interatomic distance between O and H and $\theta$ defines the angle between the $R_\text{{Rb}}$ and $R_\text{{OH}}$ vectors (see \figurename{~\ref{fig:RbOH}}). Accordingly, the geometry for $\theta=0\degree$ corresponds to the case where the Rb-O-H atoms are aligned whereas $\theta=180\degree$ corresponds to the O-H-Rb collinear configuration. The potential energy surfaces have been calculated at the coupled cluster level of theory with single, double and perturbative triple excitations (CCSD(T)) as implemented in the \begin{small}MOLPRO\end{small} program \cite{Hampel1992}. The unrestricted variant has been used for open-shell cases \cite{Knowles1993,Knowles2000}. The augmented correlation-consistent valence quintuple zeta basis set aug-cc-pV5Z (shortened AV5Z) \cite{ThomH.Dunning1988} was used for the oxygen and hydrogen atoms. We have used three different approaches to describe the rubidium atom. The first one follows the work of Byrd \textit{et al.} where the  Wood-Boring effective core potential ECP28MWB (here after abbreviated MWB) \cite{Leininger1996} was used along with the corresponding \textit{sp} functions augmented by spdf functions from the def2-QZVPP basis set \cite{Weigend2003,Weigend2005} (shortened defQZ). The final number of contracted functions in the basis is [12\textit{s}9\textit{p}4\textit{d}3\textit{f}]. In the second approach we used the more recent Dirac-Fock ECP28MDF (here after abbreviated MDF) effective core potential with the corresponding segmented \textit{spdfg} valence basis set \cite{Lim2005} which contains (13\textit{s}10\textit{p}5\textit{d}3\textit{f}1\textit{g}) functions contracted to [8\textit{s}7\textit{p}5\textit{d}3\textit{f}1\textit{g}]. Both effective core potentials (ECP) are small-core ECPs describing 28 core electrons and taking into account scalar relativistic effects (mass-velocity and Darwin terms). We also considered a third effective core potential, ECP36SDF (here after abbreviated SDF), also available for the Rb atom. The latter is a large core ECP which only leaves 1 electron of the Rb atom left for the molecular correlation treatment. The corresponding \textit{sp} functions have been used augmented by the dhf-QZVPP basis set \cite{Weigend2010}, the final number of contracted functions is [9\textit{s}8\textit{p}4\textit{d}3\textit{f}2\textit{g}]. In the three cases one set of \textit{spdf} even tempered functions of each type has been added. All available electrons, \textit{i.e} not described by the ECP, have been included in the correlation treatment. The counterpoise method was used to account for the basis set superposition error \cite{Boys1970}. This correction is likely to be important at small distances. The interatomic distance $R_\text{{OH}}$ was optimized at each $R_\text{{Rb}}$ distance at the MP2 level of theory (restricted \cite{KAAHP91} and unrestricted \citep{AAHK91} variants, as appropriate) with all available electrons correlated. 

\subsection{\label{sec:AD_PEC}Results}

\begin{figure}[h]
\includegraphics [width=8.5cm]{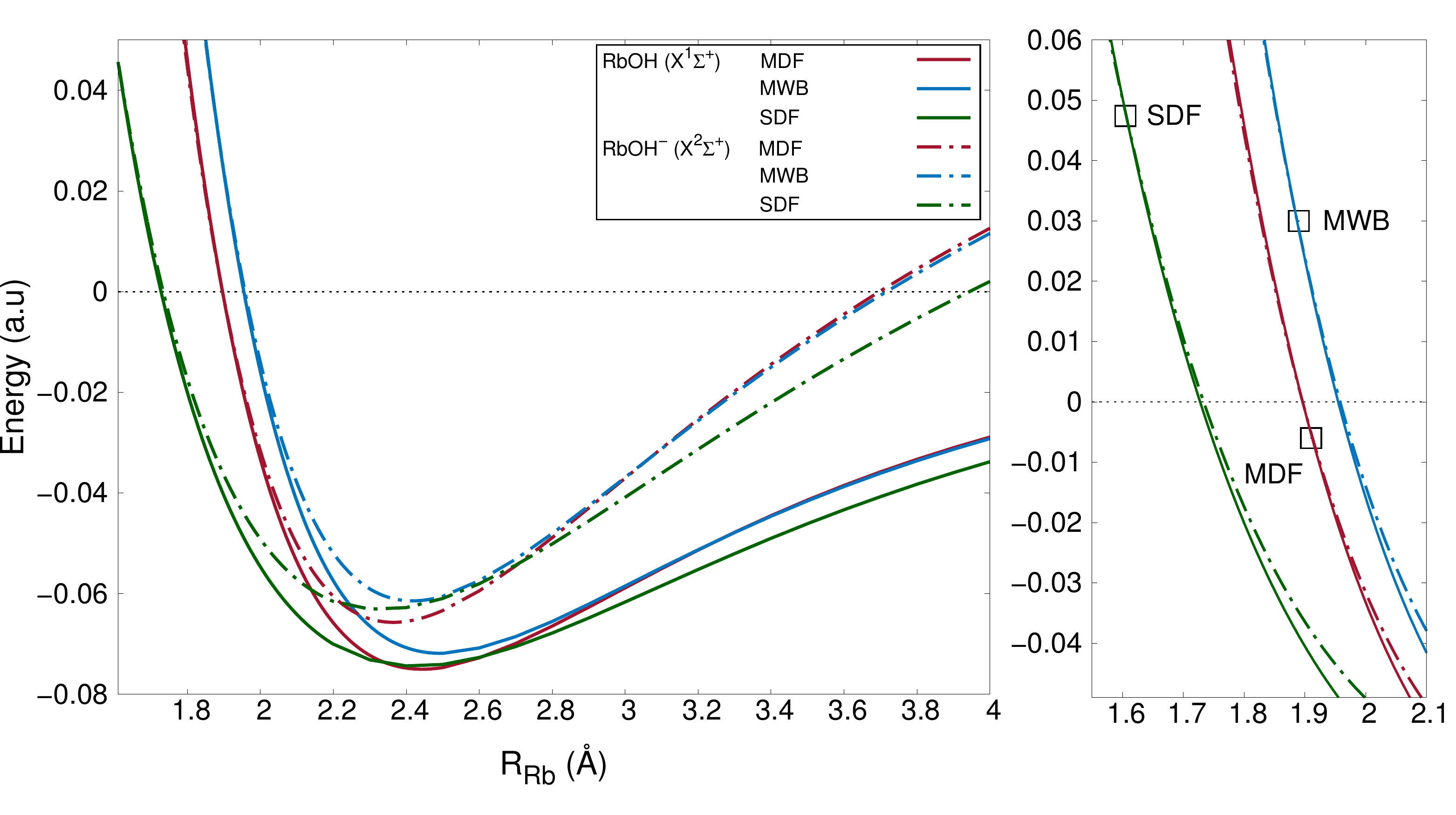}
\caption{\small PEC for the RbOH$^{-}$($^{2}\Sigma^{+}$) and RbOH($^{1}\Sigma^{+}$) molecular species for $\theta=0\degree$. The curves are calculated at the CCSD(T) level of theory using the AV5Z basis set for the O and H atoms. Results obtained using the MWB/spdfg, the MWB/defQZ and the SDF/dhf-QZ ECPs and basis set for the Rb atom are shown. The right figure shows the short-distance part of the PEC where the crossing between the anion and neutral curves are marked by squares. Solid and dashed lines stand for the anion and neutral species, respectively.}  
\label{fig:PEC_PP}
\end{figure}

\begin{figure}[h]
\begin{center}
\includegraphics[width=8.5cm]{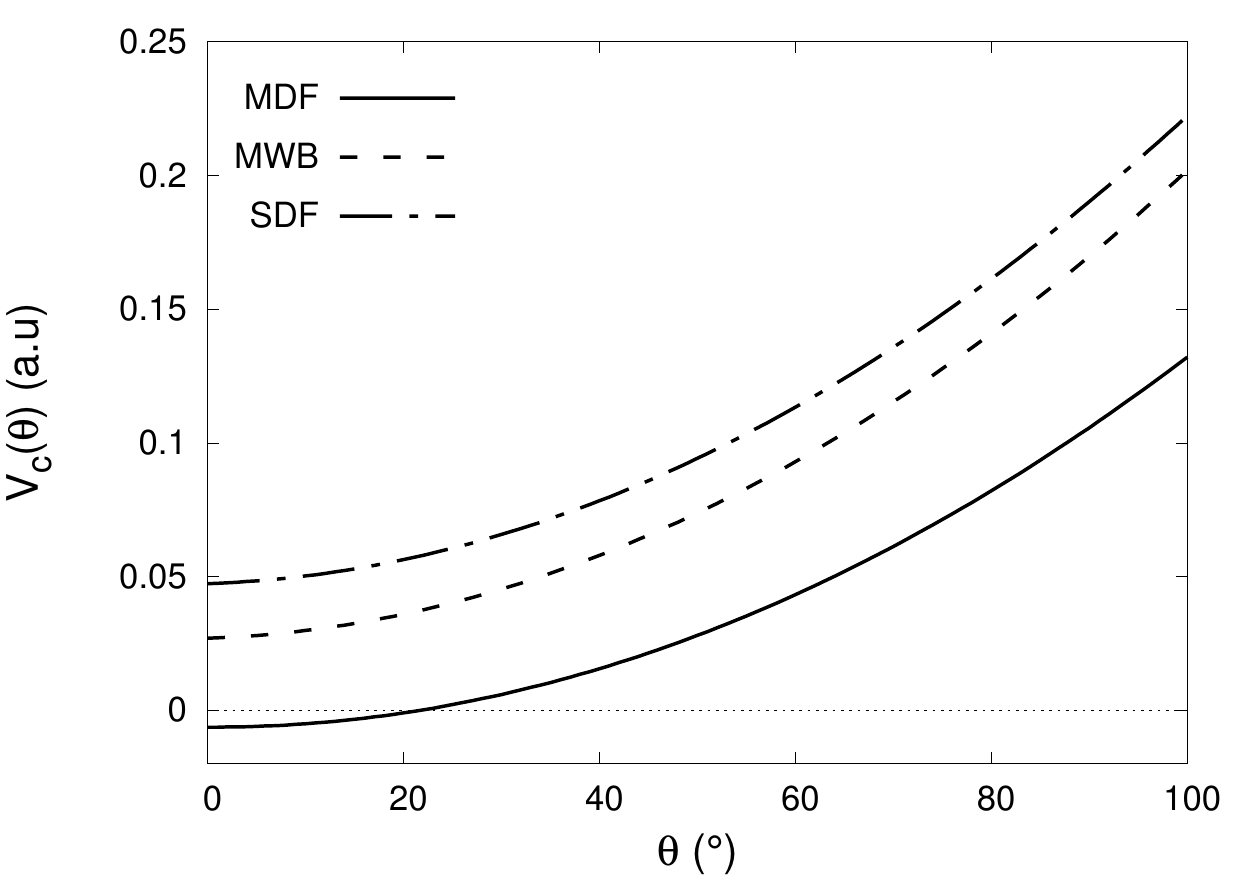}
\end{center}
\caption{\small The $V_{c}$($\theta$) function representing the energy at the crossing point between the anion (RbOH$^{-}$) and neutral (RbOH) PEC at various intramolecular angles $\theta$. The energy is taken relative to the threshold energy of the Rb($^{2}S$)+OH$^{-}$($^{1}\Sigma^{+}$) entrance channel. Results obtained using the MDF, MWB and SDF effective core potential are shown.}
\label{fig:cros_thet}
\end{figure}

The PEC at $\theta$=0 $\degree$ calculated at small $R_\text{{Rb}}$ distance for the molecular anion RbOH$^{-}$($^{2}\Sigma^{+}$) and the neutral RbOH($^{1}\Sigma^{+}$) are reported in \figurename{~\ref{fig:PEC_PP}}. A grid of 0.01 and 0.1 $ \mathring{A}$ was used at small distances (from 1.6 to 2.2 $ \mathring{A}$) and larger distances (2.2 to 4 $\mathring{\text{A}}$), respectively. The crossing points were obtained by fitting the short-distance part of the PEC with B-splines functions. Results obtained using the three different ECPs are shown, where the position of the crossing point between the neutral and anion curves is indicated by squares. The zero energy corresponds to the threshold energy of the Rb($^{2}S$)+OH$^{-}$($^{1}\Sigma^{+}$) dissociation channel calculated with each ECP. Large differences between the three ECPs are observed, in particular regarding the positions of the minimum and the crossing point. Note that the crossing height obtained using MWB is very close to the one obtained by Byrd \textit{et al.} \cite{Byrd2013} who used the same ECP and extrapolated to the complete basis set limit. 
\noindent The height of the crossing point as a function of the angle $\theta$ is shown in \figurename{~\ref{fig:cros_thet}}. The function $V_{c}$($\theta$) obtained by fitting the latter curves with a polynomial function of order 4 will be used later to model the dynamics of the associative detachment. For MDF, the crossing lies bellow the threshold energy for certain value of $\theta$ while for MWB and SDF it lies above for the entire angular space. This implies that the autodetachment region is only accessible at certain energies in the entrance channel.\\
\noindent To verify the accuracy of the ECP we have performed geometry optimization on the ground state of several diatomic molecules containing the rubidium atom. The results are given in \tablename{~\ref{RbX}}. The computed bond lengths have been compared to the available experimental values. The calculations have been performed at the CCSD(T) level of theory with all available electrons correlated. We have used different basis sets to make sure that the difference seen on \figurename{~\ref{fig:PEC_PP}} does not arise from a basis set issue.
\begin{table*}[htbp!]
\caption{\small Comparison between the calculated and experimental equilibrium distance for various molecular systems containing Rb. The geometry optimizations were performed at the CCSD(T) level of theory using different basis set and ECPs. A set of even-tempered \textit{spdf} functions has been added to all Rb basis sets.}
\begin{center}
\def\arraystretch{1.2}
\begin{tabular}{lcccccc}
\hline
\hline
  O and H basis set/Rb ECP+basis set & RbH & RbCl & RbF & RbO & RbOH & RbOH$^{-}$ \\
  \hline
  AVQZ/\textbf{MWB}\textit{sp}+def2-VQZ & 2.3891 & 2.8354 & 2.3204 & 2.3097 & 2.3405 & 2.4146  \\
  AV5Z/\textbf{MWB}\textit{sp}+def2-VQZ & 2.3682 & 2.8259 & 2.3130 & 2.2999 & 2.3309 & 2.4030 \\
  \hline
  AVQZ/\textbf{MDF}\textit{spdfg} & 2.3693 & 2.8017 & 2.2762 & 2.2600 & 2.3059 & 2.3840  \\
  AV5Z/\textbf{MDF}\textit{spdfg} & 2.3636 & 2.7962 & 2.2730 & 2.2554 & 2.3030 & 2.3800  \\
  AVQZ/\textbf{MDF}\textit{sp}+dhf-VQZ & 2.3668 & 2.7977 & 2.2776 & 2.2615 & 2.3065 & 2.3610 \\
  \hline
  AV5Z/\textbf{SDF}\textit{sp}+dhf-VQZ & 2.4439 & 2.8211 & 2.2367 & 2.2348 & 2.2769 & 2.3709 \\
  \hline
  Byrd \textit{et al.} (AVQZ/\textbf{MWB})\cite{Byrd2013} & 2.3919 & - & - & 2.3548 & 2.3408 & 2.4166 \\ 
  Exp. & 2.3667 \cite{Kato1985} & 2.7867 \cite{Rice1957} & 2.2703 \cite{Hughes1967} & 2.2541 \cite{Yamada1999} & - & - \\
  \hline
  \hline
\end{tabular}
\end{center}
\label{RbX}
\end{table*}
\noindent For RbH, the computed bond length using MWB with the AVQZ and AV5Z basis sets differs by 0.02 $ \mathring{A}$ whereas MDF results agree within 0.006 $ \mathring{A}$ and are therefore more consistent. The RbH optimized bond length obtained with MWB and a AVQZ basis set is overestimated by 0.23 $\mathring{A}$ while the MWB/AV5Z and all MDF results are close to the experimental value. Note that SDF  strongly overestimates the bond length. This is not surprising since only 2 electrons are included in the correlation treatment. For RbCl, RbF and RbO, the computed bond lengths determined with MDF and MWB differ up to 0.01  $\mathring{\text{A}}$ whereas the differences between the three ECPs are of 0.02 $ \mathring{A}$ in the best case. The MWB ECP seems to overestimate the equilibrium bond length whereas the SDF results are less consistent: overestimation for RbCl and underestimation for RbF and RbO. MDF results are much closer to the experimental values. Also shown are the converged values for the $R_\text{{RbO}}$ optimized distance for the ground state of the RbOH and RbOH$^{-}$ molecules along with the results of Byrd et al. \cite{Byrd2013}. This value represents the bond length between the Rb and the O atom. Again, MWB bond lengths are larger than MDF results for these systems. Note that in the RbOH and RbOH$^{-}$ cases, the $R_{\text{OH}}$ distance was also optimized but the result using different ECPs and basis set only varies at the third or fourth digit. The poorer results obtained with MWB can be explained by an error in the ECP for the rubidium atom, as pointed out by Weigend \textit{et al.} and confirmed by a personal communication of the authors of MWB (see \cite{Weigend2010} and reference therein). Therefore, the MWB should not be used when dealing with compounds that include the rubidium atoms. As for the SDF, the lack of electron correlation leads to inaccurate bond lengths and this ECP will not be appropriate to the description of anions.

\subsection{\label{sec:AD_Dyn}Langevin-based Dynamics} 

\begin{figure}[h]
\begin{center}
\includegraphics[width=8.5cm]{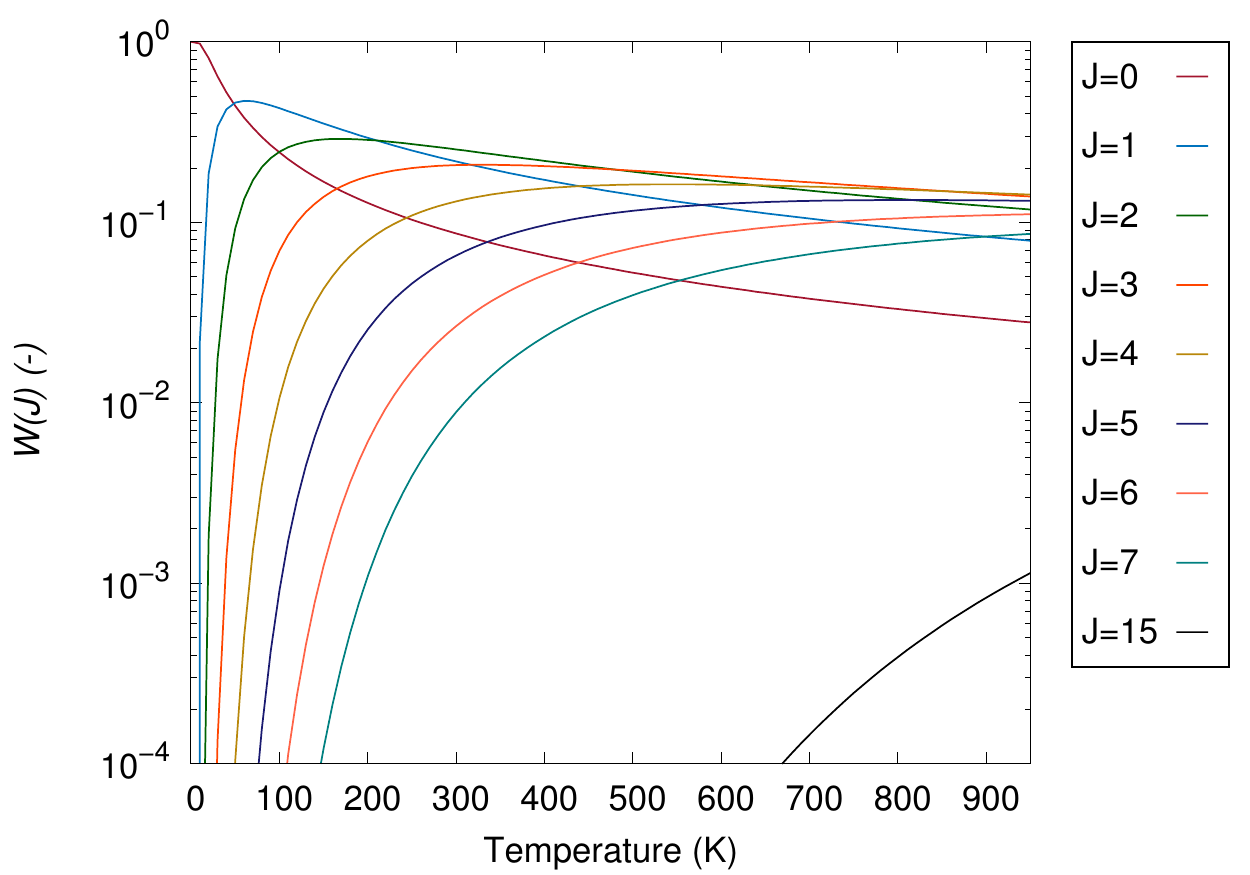}
\end{center}
\caption{\small Rotational state population for OH$^{-}$ in $v=0$.}
\label{fig:PJ}
\end{figure}

\begin{figure}[h]
\begin{center}
\includegraphics[width=8.5cm]{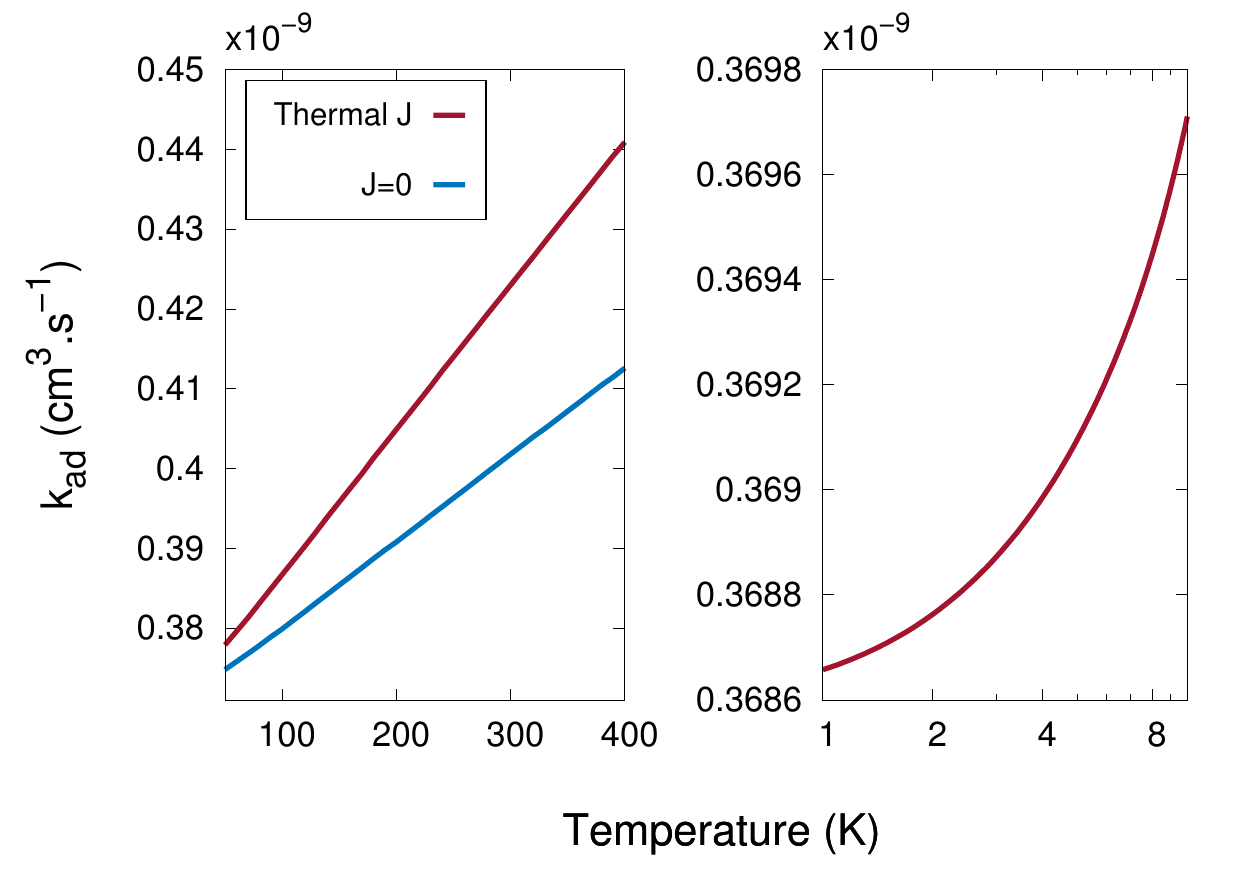}
\end{center}
\caption{\small Left: rate constant of the associative detachment reaction Rb($^{2}S$)$+$OH$^{-}$($^{1}\Sigma^{+}$)$\rightarrow \,$ RbOH($^{1}\Sigma^{+}$)$+\,e^{-}$ as a function of the temperature. The blue curve shows the results when the rotational state of OH$^{-}$ is fixed to $J$=0, the red curve shows the behaviour when the rotational state of OH$^{-}$ are thermally distributed. Right: same as left figure for smaller temperature plotted in log scale.}  
\label{fig:kad_T}
\end{figure}

\begin{figure}[h]
\begin{center}
\includegraphics[width=8.5cm]{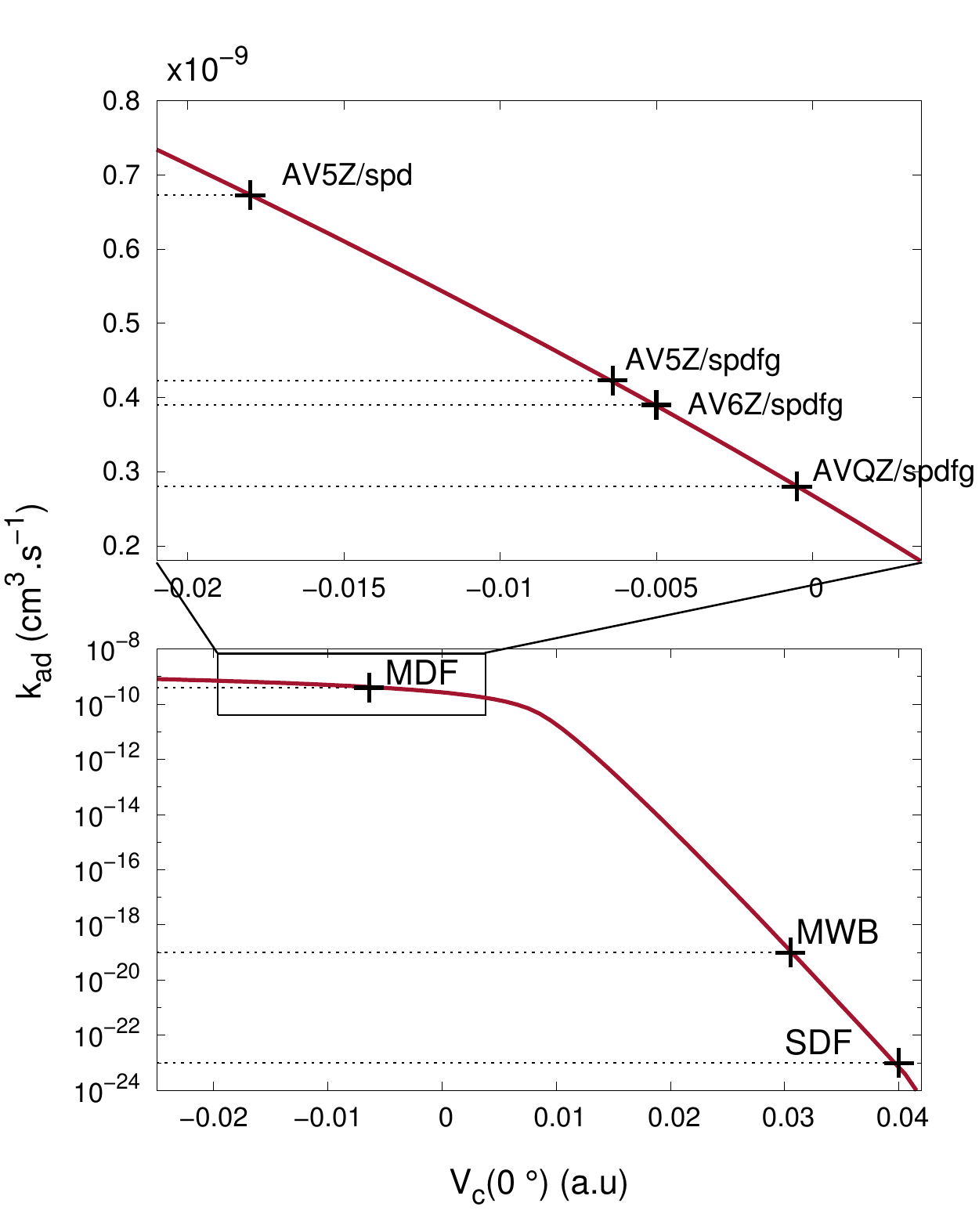}
\end{center}
\caption{\small Bottom panel: rate constant of the associative detachment reaction at 300 K as a function of the crossing point between the anion and neutral PEC at linear geometry. The calculated  crossing point at linear geometry using different ECPs are indicated by crosses. Upper panel: zoom in the MDF region. Results using different basis set with the MDF are indicated by crosses. The AVnZ basis correspond to the basis set used for the O and H atoms, for Rb the MDF with either the corresponding spdfg or spd basis has been used, in both cases extended by a set of spdf even tempered functions.}  
\label{fig:kad_y}
\end{figure}

RbOH$^{-}$ is a stable molecular anion with an electroaffinity of about 0.3 eV \cite{Byrd2013}. As has been shown in section \ref{sec:AD_PEC}, the anion and neutral PES cross in the repulsive part of the PES, allowing the anion to enter into the autodetachment region. In the present case, it is the collisional kinetic energy which allows the electron to be ejected via a surface crossing mechanism \cite{Simons1998}. In order to calculate the rate constant for the associative detachment reaction we have used a Langevin model similar to the one described by Byrd et al. The assumptions used in the model are the following: (i) the transition probability from the anionic to the neutral state is 0 for $R_{\text{Rb}}>R_{c}$ and 1 for $R\leq R_{c}$ where $R_{c}$ is the nuclear position at the crossing point, (ii) the energy of the colliding partners follows a Maxwell-Boltzmann distribution and (iii) the OH$^{-}$ molecules are rotating sufficiently fast as to average the collisions over the angular space. Langevin models have been successfully compared to full quantum scattering calculations and to experimental results for different associative detachment reactions (O$^{-}$+H \citep{Acharya1985}, S$^{-}+$H  \cite{Fedchak1993}, H+Cl$^{-}$ \cite{Cizek1999}, H+Br$^{-}$ and H+F$^{-}$ \cite{Cizek2001,Domcke1985}, CO+O$^{-}$ and H$_{2}$+O$^{-}$ \cite{Schultz1973}). In our case the treatment is simplified since only one molecular state of the anion RbOH$^{-}$ correlates to the entrance channel, no barrier is present along the PEC, and no other reaction channels are available in the energy range of interest. The use of the Langevin cross section should therefore be a reasonable choice.\\
We summarize hereafter the main equations that lead to the rate constant, for a more detailed description see \cite{Byrd2013}. The associative reaction only occurs when the crossing point is reached. This depends on the crossing height $V_{c}$($\theta$) and the energy in the Rb($^{2}S$)+OH$^{-}$($^{2}\Pi$) entrance channel: $\varepsilon$ + $T$($v$,$J$) where $\varepsilon$ is the collision energy and $T$($v$,$J$) the ro-vibrational energy of OH$^{-}$. The total cross section can be defined as:
\begin{eqnarray}
\sigma_{tot}(v,J,\varepsilon)=\rho_{c}(v,J,\varepsilon)\,\sigma_{L}(\varepsilon)
\label{sigtot}
\end{eqnarray}
where,
\begin{eqnarray}
\rho_{c}(v,J,\varepsilon)&=&\frac{1}{2}\,\int^{\pi}_{0}\Xi(\varepsilon+T(v,J)-V_{c}(\theta))\,\sin\theta\,d\theta \\
&=&\frac{1}{2}\,\Big(1-\cos(\theta_{max}(v,J,\varepsilon))\Big)
\label{rhoc}
\end{eqnarray}
is the accessible angular space, and
\begin{eqnarray}
\sigma_{L}(\varepsilon)=\pi\sqrt{\frac{2\,\alpha_{d}}{\varepsilon}}
\label{sigL}
\end{eqnarray}
\noindent is the dipole polarization Langevin cross section. $\alpha _{d}$ is the dipole polarizability of Rb($^{2}S$) which is 318.6 a.u \cite{Derevianko1999}. The $\Xi(\varepsilon,V_{c}(\theta),T(v,J))$ function is the Heavyside function used in \cite{Byrd2013}. The $\theta _{max}$ is the angle above which $V_{c}(\theta)\geq\varepsilon$ + $T$($v$,$J$). The value of $\theta _{max}$ thus depends on $v$, $J$ and $\varepsilon$. The rotational constant, vibrational frequency and coupling terms used to obtain $T$($v$,$J$) are taken from \cite{Rosenbaum1986}. The rate constant for associative detachment can be written as
\begin{equation}
k_{ad}(v,T)=\sum_{J=0}^{\infty}\Big(\,W(J)\int_{0}^{\infty}f(\varepsilon)\,\sigma_{tot}(v,J,\varepsilon)\,d\varepsilon\Big)
\label{kad}
\end{equation}
and
\begin{eqnarray}
W(J)=\frac{1}{Q_{rot}}\,(2J+1)\,e^{(\frac{-E_{J}}{k_{b}T})}
\label{W(J)}
\end{eqnarray}
\noindent where $f$($\varepsilon$) is the Maxwell-Boltzmann distribution. The $W$($J$) terms account for the weight of each rotational state of OH$^{-}$ where the factor (2$J$+1) is the degeneracy factor and $Q_{rot}$ is the rotational partition function. $W$($J$), which represent the rotational state population of OH$^{-}$, can be seen in \figurename{~\ref{fig:PJ}} as a function of the temperature. In equation \ref{kad}, summing up to $J=15$ was found sufficient since $W$($J$) becomes very small for larger J values. 
\noindent Using the $V_{c}$($\theta$) function obtained with the MDF calculations and assuming that the OH$^{-}$ ions are in their vibrational ground state (thus $T$($v$,$J$)=$T$($0$,$J$)) we obtained a rate constant of $4\times10^{-10}$ cm$^{3}$s$^{-1}$ at 300 K. This result agrees well with the first experimental results of Deiglmayer \textit{et al.} \cite{Deiglmayr2012} where the measured rate constant was 2$^{+2}_{-1}\times10^{-10}$cm$^{3}$s$^{-1}$ for a temperature between 200 K and 600 K.\\
We have also calculated the rate constant of the associative detachment $k_{ad}$ as a function of the temperature. The resulting plots can be seen in \figurename{~\ref{fig:kad_T}}. As the temperature decreases, the rate constant decreases. Two effects contribute to the latter : the decrease of the contribution of large $J$ values in $W$(J) and of the collision energy $\varepsilon$. Both contributions reduce the accessible angular space, \textit{i.e} the value of $\theta_{max}$, in equation (\ref{rhoc}), and thus the cross section $\sigma_{tot}$ in equation (\ref{kad}). The rate constant $k_{ad}$ only changes slightly with temperature, e.g $k_{ad}$ only decreases from a factor 1.2 between 400 and 2 K. The rotational state of OH$^{-}$ may not be in thermal equilibrium in the hybrid trap where the reaction takes place. A recent theoretical study \cite{Gonzalez-Sanchez2015} showed that the $J$=0 state of OH$^{-}$ may actually be the most populated one in the trap. Therefore, we have also shown the calculated rate constant for $J$=0 fixed at all temperature. However, this only affects slightly the rate constant for high temperature \textit{e.g} at 300 K the difference is only of 6\%. \\
On the other hand, the rate constant strongly depends on the height of the crossing point between the PESs of the neutral and the anion. Since the crossing is located in the repulsive region of the PES, its position is very sensitive to the computational method and basis set used. To illustrate this dependence, we have represented in \figurename{~\ref{fig:kad_y}} the rate constant at 300 K as a function of the crossing height at linear geometry. The rate constant was computed using the function V$_{c}(\theta)$ obtained with the MDF ECP (see \figurename{~\ref{fig:cros_thet}}) by varying $V_c(0\degree)$.
Two regimes can be distinguished. When the crossing point $V_c(0\degree)$ is located below the entrance channel, the rate shows an almost linear dependence on the crossing height. On the other hand, when $V_c(0\degree)$ is located above the entrance channel, the rate decreases exponentially with increasing crossing height.
\noindent The rates obtained based on calculations performed with the MDF ECP and various basis sets are shown on the upper panel in \figurename{~\ref{fig:kad_y}}. The crossing point at linear geometry is always located below the entrance channel and the resulting rate is comprised between $3\times 10^{-10}$ and $7\times 10^{-10}$ cm$^{-3}$s$^{-1}$, which allows us to define a theoretical uncertainty in the framework of the Langevin model. The most accurate results are expected to be obtained for the largest basis sets, i.e. AV5Z/spdfg and AV6Z/spdfg. It should be noted that the basis set superposition error does not have a significant effect on the crossing point since the corrections to the anion and neutral PESs are similar for large basis sets. As discussed in section \ref{sec:AD_PEC}, the calculations performed with the MWB and SDF ECPs lead to a crossing point above the entrance channel. This results in an associative detachment rate constant that is several orders of magnitude smaller than the MDF and experimental values. This confirms the fact that the MWB ECP should not be trusted for small values of $R_{\textrm{Rb}}$ whereas the lack of correlated electrons in the SDF calculations leads to a bad description of the chemical bond, especially for the anion.\\
To conclude, we can say that the rate constant of the associative detachment reaction Rb($^{2}S$)$+$OH$^{-}$($^{1}\Sigma^{+}$)$\rightarrow \,$ RbOH($^{1}\Sigma^{+}$)$+\,e^{-}$ is almost constant with respect to the temperature, i.e the collision energy $\varepsilon$ and the rotational population distribution of OH$^{-}$. In contrast, the rate strongly depends on the position of the crossing point which lays in the repulsive region and is therefore difficult to compute accurately. However, compare to the experimental results, we obtain agreement using the MDF ECP with large basis sets. It is worth mentioning that this problem would also appear if a full quantum description is used instead of the Langevin model to describe the associative detachment reaction since the results will also depend on the PES and on the position of the crossing point.          

\section{\label{sec:Exc}Collision with excited rubidium}

Collisions between electronically excited rubidium and OH$^{-}$ are also likely to occur in the co-trapping experiment. Moreover, the amount of rubidium in its first excited state can be tuned by varying the intensity of the laser used in the magneto optical trap \cite{Deiglmayr2012}. Charge transfer and associative detachment reactions could both occur from the excited entrance channel, since they are both exothermic. Collision between molecular ions and ultracold Rb have already been studied and have shown some interesting features \cite{Hall2012}. To investigate these possibilities, we have calculated the PESs involving the excited reaction channels.

\subsection{\label{sec:Exc_method}Computational method}

In order to calculate the different PESs of the RbOH$^{-}$ molecular system we have used the internally contracted multi-reference configuration interaction method (ic-MRCI) \cite{Knowles1992} as implemented in the \begin{small}MOLPRO\end{small} program. The reference wave function on which the single and double excitations are performed is a state-averaged complete active space wave function (SA-CASSCF) [cite] with an active space covering 6$\sigma$ and 3$\pi$ molecular orbitals. The first $\sigma$ orbital, which corresponds to the 1$s$ orbital of the oxygen atom was kept frozen, i.e taken from a previous Hartree-Fock calculation, to avoid rotation between the 1$s_\text{{O}}$ and 4$s_\text{{Rb}}$ orbitals. The corresponding $C_{s}$ orbitals where used for the non-linear cases, hence 10$a'$ and 3$a''$ orbitals. We have included all states that correlate to the first 3 dissociation channels, \textit{i.e} Rb+OH$^{-}$, Rb$^{-}$+OH and Rb*+OH$^{-}$, in the state-average procedure. The AVTZ basis set was used to describe the O and H atoms \cite{ThomH.Dunning1988}. For the rubidium atom we used the MDF ECP with the corresponding \textit{spdfg} valence basis set \cite{Lim2005} and a set of \textit{spdf} even tempered functions. The OH interatomic distance, $R_\text{{OH}}$, was kept fixed at the OH$^{-}$ experimental value of 0.9643 $\mathring{\text{A}}$ \cite{Rosenbaum1986}. We have included the Davidson correction using rotated reference energies to account for the size inconsistency problem \cite{Langhoff1974,Meissner1988,Werner2008}. The numbering of the different states used in \figurename{~\ref{fig:PEC_tot}} and \tablename{~\ref{tabsym}} has been kept for the following results.  

\subsection{\label{sec:Exc_result}Potential energy surfaces}

\begin{figure}[h]
\begin{center}
\includegraphics[width=8.5cm]{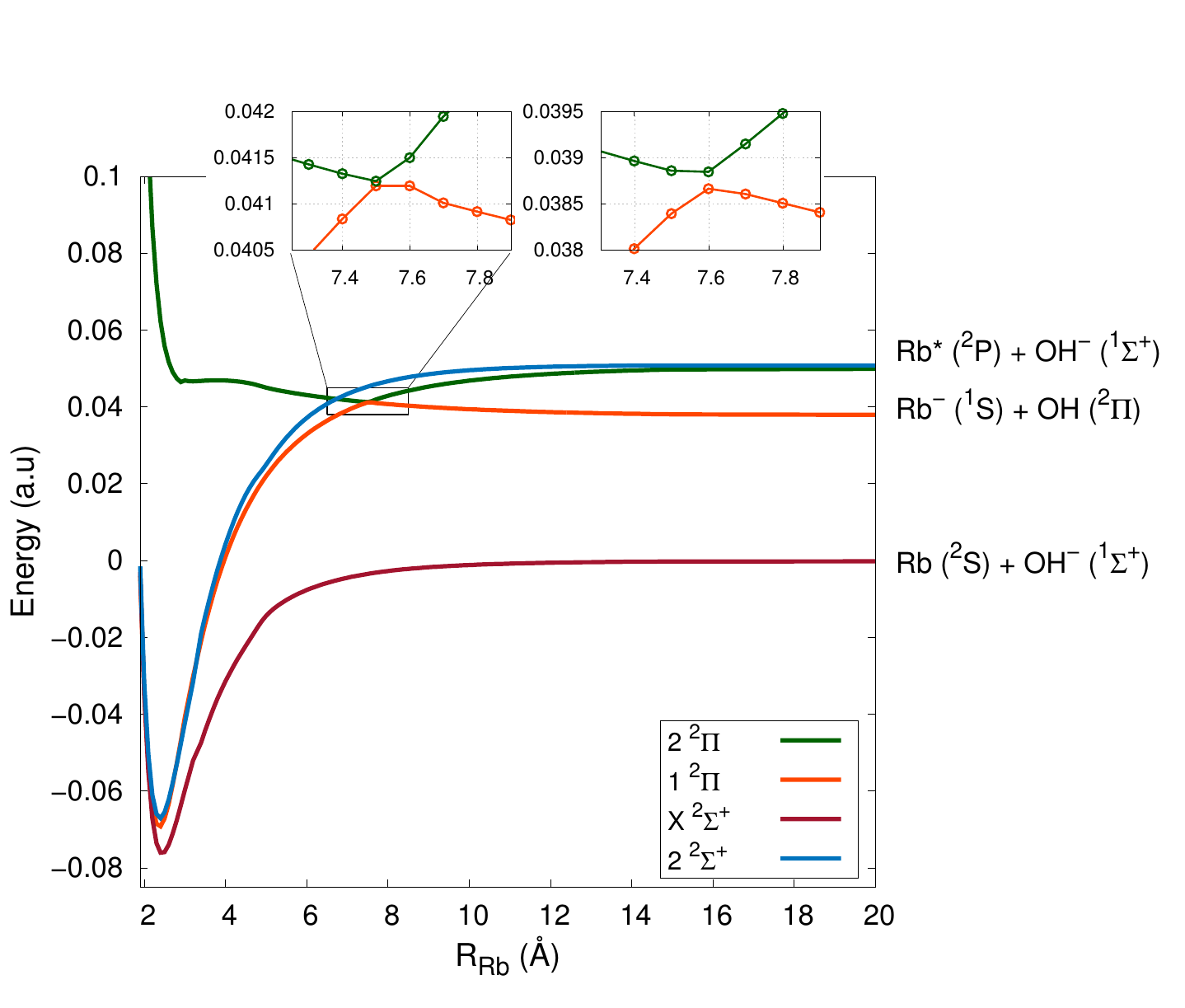}
\end{center}
\caption{\small PECs for the two first $^{2}\Sigma$ states and the two first $^{2}\Pi$ states of the RbOH$^{-}$ anion at linear geometry ($\theta=0 \degree$). The dissociation states are depicted. The left and right insets show a zoom of the avoided crossing with and without the Davidson correction, respectively.}    
\label{fig:PEC_0}
\end{figure}

\begin{figure}[h]
\begin{center}
\includegraphics[width=8.5cm]{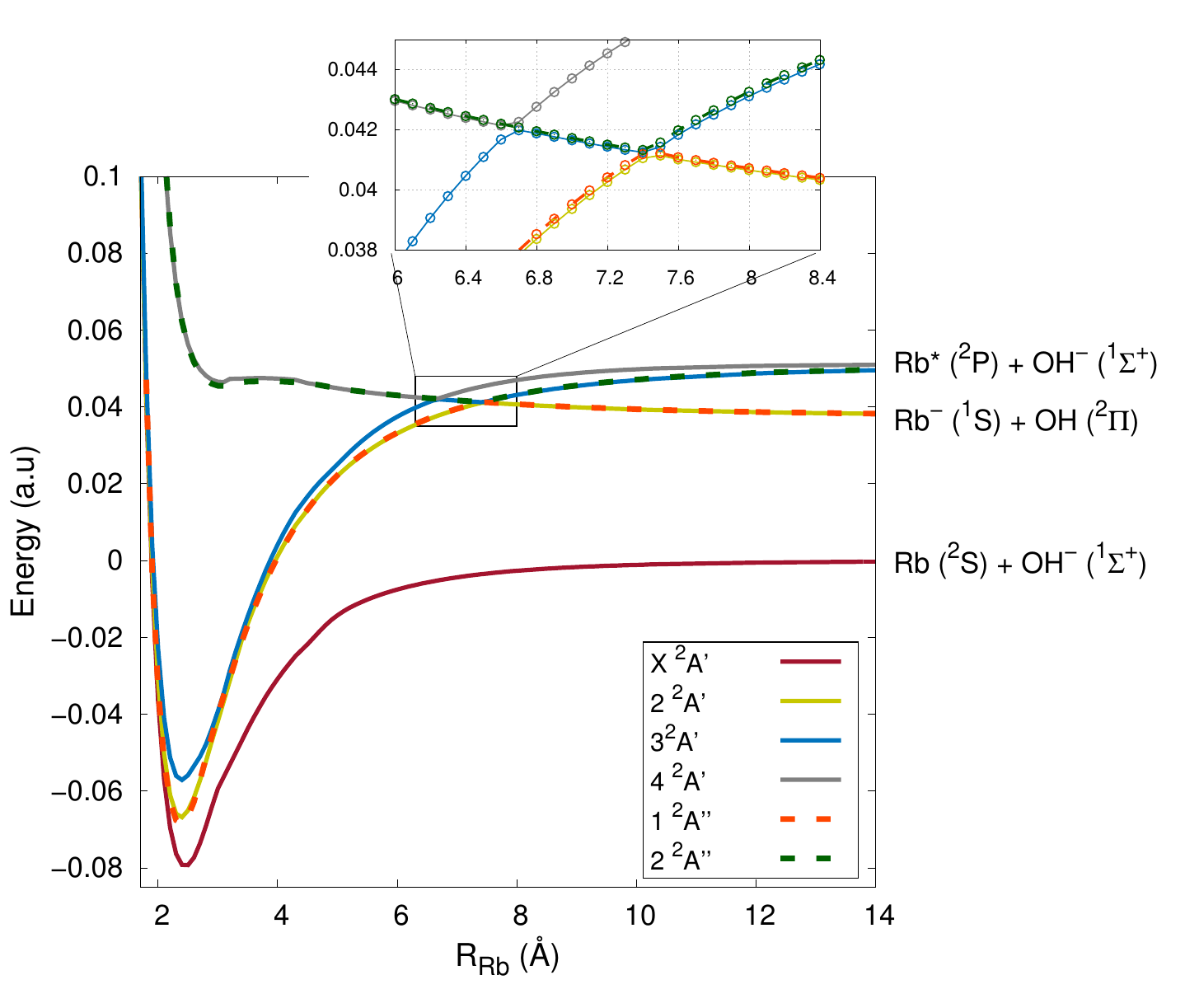}
\end{center}
\caption{\small PEC for the 4 first $^{2}A'$ and the two first $^{2}A''$ states of the RbOH$^{-}$ anion at bent geometry ($\theta$ =20 $\degree$). The dissociations states are depicted. The zoomed-in figure shows the different avoided crossings.}
\label{fig:PEC_20}
\end{figure}

\begin{figure}[h!]
\begin{center}
\includegraphics[width=8.5cm]{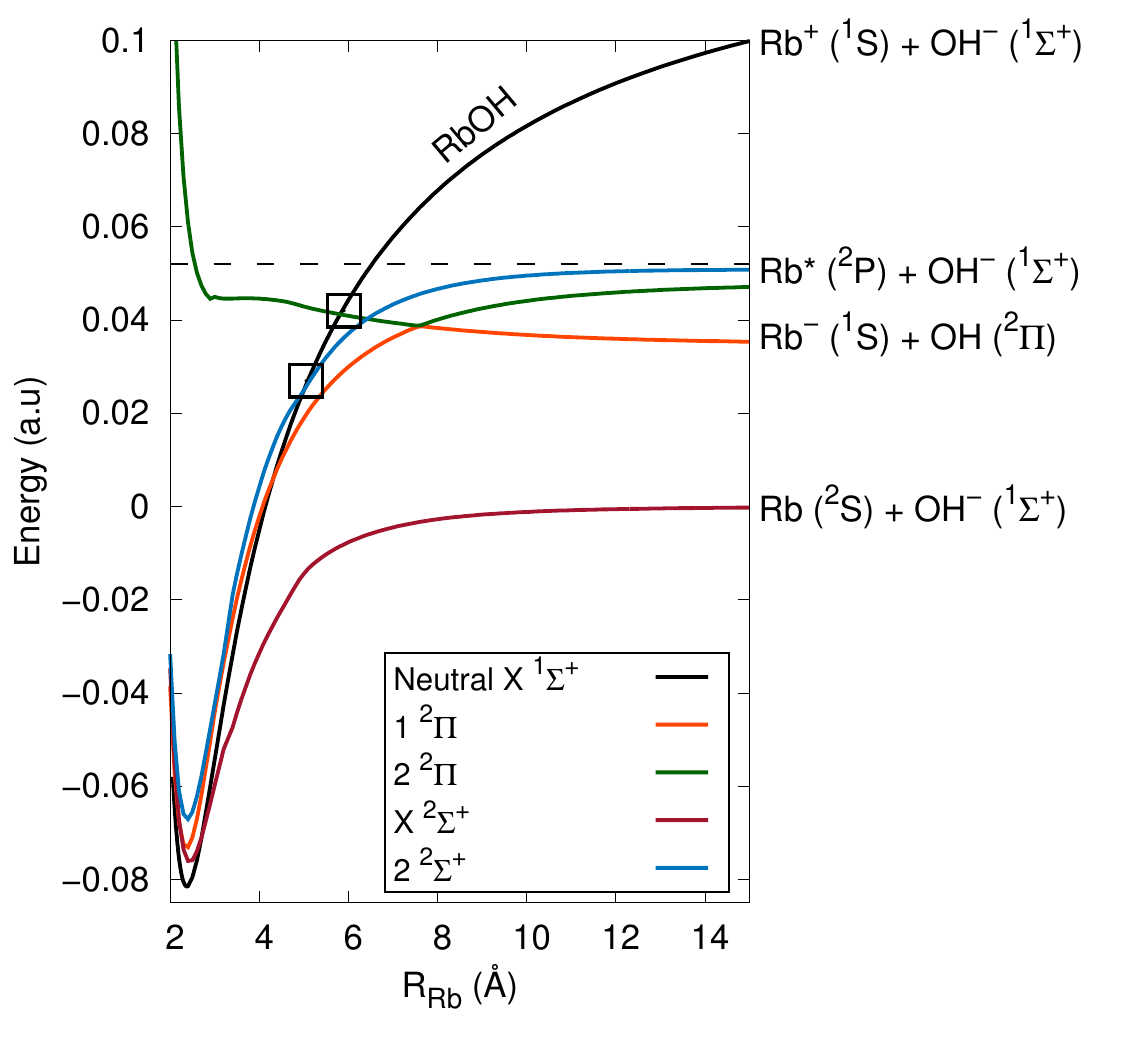}
\end{center}
\caption{\small PECs for the 4 first $^{2}A'$ and the two first $^{2}A''$ states of the RbOH$^{-}$ anion (coloured curves) and the ground state of the neutral RbOH specie (black curve) at linear geometry. The dissociation states are depicted. The crossing points between anions and neutral curves are indicated by squares. These crossings define the entrance of the autodetachment region.}
\label{fig:PEC_0_N}
\end{figure}

\begin{figure}[h!]
\begin{center}
\includegraphics[width=8.5cm]{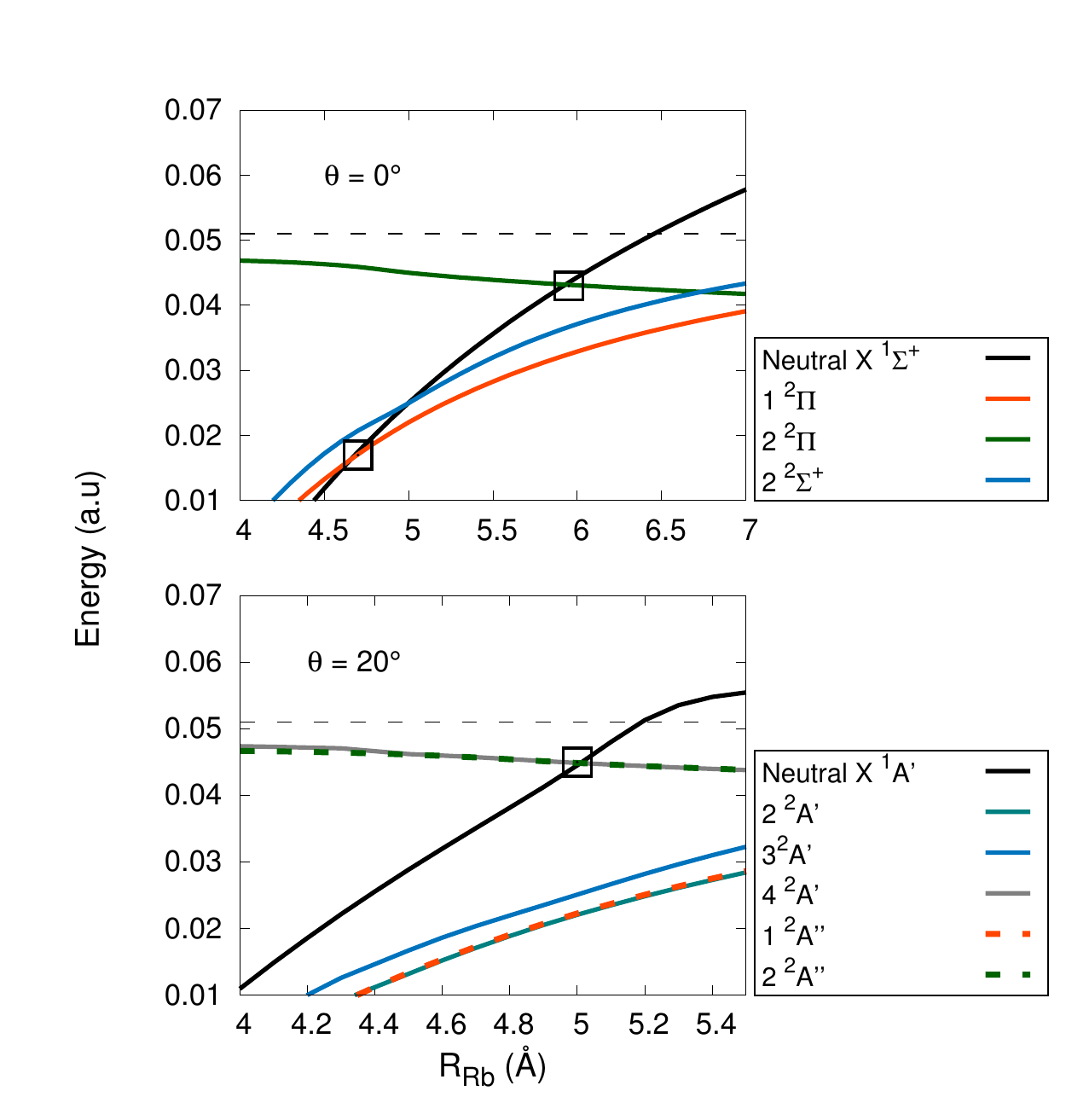}
\end{center}
\caption{\small Zoom in the crossing region between the excited anion curves and the neutral ground state curves at linear geometry (upper panel) and 20 $\degree$ (bottom panel). The crossings are indicated by squares. The dashed lines represents the energy of the Rb($^{2}P$)+$\text{OH}^{-}$($^{1}\Sigma ^{+}$) excited entrance channel. Energies are calculated relative to ground state Rb($^{2}S$)+OH$^{-}$($^{1}\Sigma ^{+}$) channel.}
\label{fig:cros_exc}
\end{figure}

The calculated adiabatic PECs at linear geometry are shown on \figurename{~\ref{fig:PEC_0}}. One can see an avoided crossing, indicated by a square between the two $^{2}\Pi$ states (labelled $1\,^{2}\Pi$ and $2\,^{2}\Pi$, respectively) that allows the charge transfer reaction Rb$(^{2}P)+$OH$^{-}(^{1}\Sigma ^{+}) \rightarrow \,\, $Rb$^{-}(^{1}S)+$OH$(^{2}\Pi)$ to occur via non adiabatic coupling terms. The insets in \figurename{~\ref{fig:PEC_0}} show a zoom of the avoided crossing with and without Davidson correction. The energy gap becomes smaller when the correction is taken into account. This arise from the difference in the Davidson correction for both states, where the negative charge is either located on the Rb or the O atom.  
At bent geometries, the two $^{2}\Pi$ states split into two $^{2}A'$ and two $^{2}A''$ states which undergo avoided crossings with each other and with the $^{2}A'$ state arising from the $2\,^{2}\Sigma ^{+}$ state. This is shown on the inset in \figurename{~\ref{fig:PEC_20}} along with the PECs for $\theta$=20 $\degree$. We thus have 3 avoided crossings where two are cuts through a conical intersection arising from a pseudo Jahn-Teller effect \cite{Jahn1937,Pearson1975}.
\noindent The excited rubidium is present in its $J=3/2$ fine structure state in the MOT, the collision will therefore follow the $2\,^{2}\Pi$ PEC. Taking into account the fine structure of Rubidum, which arises from spin-orbit coupling, the charge transfer reaction Rb$(^{2}P_{3/2})+$OH$^{-}(^{1}\Sigma ^{+}) \rightarrow \, $Rb$^{-}(^{1}S)+$OH$(^{2}\Pi)$ is exothermic with an energy release of 0.241 eV. This value represents the energy difference between the entrance and exit channels and is obtained by subtracting the electron affinity of Rb (0.496 eV \cite{Frey1978}) from the electron affinity of OH (1.8290 eV \cite{Smith1997}) and then subtracting the obtained value by the excitation energy of the Rb($^{2}P_{3/2}$) state (1.589 eV \cite{Sansonetti2006}). When using the calculated energy at 1000 $\mathring{\text{A}}$ we obtain 0.368 eV, which is 0.127 eV larger that the experimental value. Several factors contribute to the discrepancy between the calculated and experimental values. The first is the omission of the spin-orbit splitting in our calculations. The second is the fact that two electron affinities are involved, which is known to be difficult to calculate accurately by quantum chemistry methods \cite{Simons2008a,Rienstra-Kiracofe2002,Feller1989}. To illustrate these difficulties, we have computed the electron affinity of OH and Rb and the excitation energy of Rb at the CASSCF/ic-MRCI level of theory using different basis sets and active spaces. We have also tested the effect of the Davidson correction. The results are depicted on \tablename{~\ref{Rb} and \ref{OH}} for Rb and OH, respectively.\\
From \tablename{~\ref{Rb}} we observe that the Davidson correction is zero for the calculated EA and $\bigtriangleup E$ when the 4s and 4p orbitals are closed. There is indeed only one and two correlated electron for the neutral and anionic species, respectively. When the 4\textit{s} and 4\textit{p} orbitals are open and the corresponding orbitals correlated, the Davisdon correction significantly improves the results. A clear trend can be seen for the excitation energies ($\bigtriangleup E$), which converges towards the experimental result with increasing active space size. Unfortunately, while including the 4\textit{s} and 4\textit{p} orbitals improves the excitation energy, the EA is worsened. One possible explanation is that the core-core and core-valence correlation energy is more important for the neutral than for the anion since the orbitals of the latter are somewhat more diffuse. The contribution for the neutral and anion are not correctly balanced and the EA will in consequence become smaller when including the 4\textit{s} and 4\textit{p} orbitals. The third columns shows the results using an active space corresponding to the one used in the molecular case. The EA is overestimated by 0.025 eV and the excitation energy underestimated by 0.113 eV. Comparison with results obtained from separated CASSCF wave function for the neutral and anion show than the deviations are mostly due to the inclusion of the anion in the state average procedure, which destabilizes the neutral orbitals. Increasing the active space to include the 4d and 6s orbitals of Rb would improve the results, however this become untreatable at the molecular level from a computational point of view. These results highlight the very well known difficulty to correctly describe anions.\\
\begin{table}[!htbp]
\caption{\small Electron affinity and energy of the first excited state of rubidium, calculated at the CASSCF and icMRCI level of theory with different active spaces. The MDF ECP and the corresponding spdfg basis set augmented by a set of \textit{spdf} even tempered functions has been used. The SA-CASSCF reference wave function is state averaged over the $^{2}S$ and $^{2}P$ states of Rb and the $^{1}S$ of the Rb$^{-}$. The results obtained with and without the Davidson correction are labelled D and nD, respectively. CCSD(T) results for the EA using the same basis set: -0.474 eV and -0.481 eV without and with the correlation of the 4\textit{s} and 4\textit{p} orbitals, respectively. Experimental value: EA=-0.496 eV \cite{Frey1978}, $\bigtriangleup E_{J=1/2}$=1.559 eV \cite{Sansonetti2006} and $\bigtriangleup E_{J=3/2}$=1.589 eV \cite{Sansonetti2006}.}
\begin{center}
\def\arraystretch{1.3}
\resizebox{\columnwidth}{!}{
\begin{tabular}{llcccc}
\hline
\hline
& & \multicolumn{4}{c}{Active atomic orbitals}\\
\cline{3-6}
& & 5\textit{s},5\textit{p} & 5\textit{s},5\textit{p},6\textit{s},4\textit{d} & 4\textit{s},4\textit{p},5\textit{s},5\textit{p} & 4\textit{s},4\textit{p},5\textit{s},5\textit{p},4\textit{d},6\textit{s}\\
\hline
\multirow{3}{*}{EA (eV)}& CASSCF   & -0.476 & -0.468 & -0.469 & -0.459  \\
                        & MRCI-nD & -0.471 & -0.471 & -0.544 & -0.464  \\
                        & MRCI-D  & -0.471 & -0.471 & -0.521 & -0.470  \\
\hline
\multirow{3}{*}{$\bigtriangleup E$ (eV)} & CASSCF   & 1.245 & 1.348 & 1.252 & 1.39  \\
                                         & MRCI-nD & 1.348 & 1.348 & 1.402 & 1.555  \\
                                         & MRCI-D  & 1.348 & 1.348 & 1.476 & 1.574  \\
  \hline
  \hline
\end{tabular}
}
\end{center}
\label{Rb}
\end{table}
The results obtained for OH in \tablename{~\ref{OH}} show a much stronger dependence on the size of the active space, Davidson correction and basis set. The results converge towards the experimental value for large active space and basis set. The electron affinity calculated with the AVTZ basis set with an active space covering the valence atomic orbitals is 0.403 eV above the CCSD(T)/AVQZ results. The Davidson correction improves the result by 0.348 eV. This is not surprising since the correction accounts partially for the quadruple excitation terms, known to be important in OH$^{-}$ \cite{Martin2001}.\\
\noindent Hence, the discrepancy between the calculated electron affinities and excitation energy and their respective experimental value explains the deviation between the calculated and experimental energies at the dissociation limit of the RbOH$^{-}$ specie.\\
\begin{table*}[htbp!]
\caption{\small Electron affinity of OH, calculated at the CASSCF-icMRCI level of theory using different active space and basis set. The SA-CASSCF reference wave function is state averaged over the anionic OH$^{-}(^{1}\Sigma^{+})$ and neutral OH($^{2}\Pi$) states. The $1\sigma$ molecular orbital corresponding to the $1s_{\text{O}}$ atomic orbital was kept frozen. The R$_\text{{OH}}$ distance was held fixed at 0.9643 $\mathring{\text{A}}$. The results obtained with and without the Davidson correction are labelled D and nD, respectively. CCSD(T)/AVQZ results: EA=-1.796 eV. Experimental value: EA=-1.828 eV \cite{Smith1997}.}
\begin{center}
\begin{tabular}{lccccccccccc}
\hline
\hline
& \multicolumn{11}{c}{Basis set}\\
\cline{2-12}
&\multicolumn{3}{c}{AVDZ} & & \multicolumn{3}{c}{AVTZ} & & \multicolumn{3}{c}{AVQZ}  \\
\cline{2-4} \cline{6-8} \cline{9-12}
 Active orbitals & CASSCF & nD & D & & CASSCF & nD & D & & CASSCF & nD & D \\
 \hline
  $1s_{\text{H}}$, $2s_\text{{O}}$, $2p_\text{{O}}$  & -0.065  & -1.393 & -1.741 & & -0.052 & -1.405 & -1.742 & & -0.048 & -1.437 &-1.766 \\
  $1s_\text{{H}}$, $2s_\text{{O}}$, $2p_\text{{O}}$, $3s_\text{{O}}$ & -1.274 & -1.603 &-1.631 & & -1.254 & -1.676 &-1.718 & & -1.253 & -1.723 & -1.768  \\
  $1s_\text{{H}}$, $2s_\text{{O}}$, $2p_\text{{O}}$, $3s_\text{{O}}$, $3p_\text{{O}}$  & -1.301 & -1.603 &-1.637 & & -1.287 & -1.678 & -1.724 & & -1.286 & -1.724 & -1.774 \\
  $1s_\text{{H}}$, $2s_\text{{O}}$, $2p_\text{{O}}$, $3s_\text{{O}}$, $3p_\text{{O}}$, $4s_\text{{O}}$ & -1.300 & -1.604 &-1.638 & & -1.281 & -1.678 & -1.724 & & -1.280 & -1.723 & -1.773 \\
\hline
\hline
\end{tabular} 
\end{center}
\label{OH}
\end{table*}
Even if the presented results show some lack of accuracy, we can already extract some trends concerning the reactions occurring from the excited entrance channels. The associative detachment reaction Rb$(^{2}P_{3/2})+$OH$^{-}(^{1}\Sigma ^{+}) \rightarrow \, $RbOH$(^{1}\Sigma ^{+})+e^{-}$ is energetically possible, as shown in \figurename{~\ref{fig:PEC_0_N}} where the neutral RbOH($^{1}\Sigma ^{+}$) curve along with those corresponding to the ground and excited states of RbOH$^{-}$ are plotted. The anion curve enters the autodetachment region around $R_{Rb}$=5 $\mathring{\text{A}}$, below the threshold energy of the Rb($^{2}P_{\frac{3}{2}}$)+OH$^{-}$($^{1}\Sigma ^{+}$) entrance channel (dashed black line in \figurename{~\ref{fig:PEC_0_N}}) and after the avoided crossing between the two $^{2}\Pi$ states. Note that the collisional detachment reaction $Rb(^{2}P_{\frac{3}{2}})+$OH$^{-}(^{1}\Sigma ^{+}) \rightarrow \, $Rb$(^{2}S)+$OH$(^{2}\Pi)+e^{-}$ is not accessible in the low temperature regime since the exit channel is around 0.2 eV above the entrance channel. A zoom in the crossing region is shown in \figurename{~\ref{fig:cros_exc}} for $\theta$=0$\degree$ and $\theta$=20$\degree$. The neutral PECs have been obtained using the same active space, ECP and basis set used for the anion.\\
In the following discussion on the dynamics we limit ourselves to the linear case for simplicity. The charge transfer and the associative detachment reaction will compete with each other. Assuming a transition probability close to unity from the anionic state to the neutral state when entering the autodetachment region, the AD reaction will prevail over the CT reaction. Indeed, the wave packet representing the Rb($^{2}P$)+OH$^{-}$($^{1}\Sigma^{+}$) molecular system  will enter from the green curve in \figurename{~\ref{fig:PEC_0_N}} and \ref{fig:PEC_0}, reach the avoided crossing and either adiabatically follow the green curve or diabatically cross over the orange curve. In both cases the wave packet will enter the autodetachment region and undergo associative detachment with a probability close to unity, preventing the possibility for the wave packet to exit via the CT channel. This assumption is very rough and a full quantum description would be necessary to get a correct branching ration between the AD and CT reactions. Moreover, the crossing between the neutral and excited curves will vary with $\theta$ and could lie above the energy at dissociation for certain values of $\theta$ as in the case of the ground state (see section \ref{sec:AD_PEC}). This can already be seen in \figurename{~\ref{fig:cros_exc}} where the crossing at $\theta$=20$\degree$ occurs at a smaller distance and slightly higher energy than for $\theta$=0$\degree$. In addition, the barrier which is present around 4 $\mathring{\text{A}}$ along the $2\,^2\Pi$ (2\textit{A'}) curve will also contribute to the dynamic. These two effects will probably increase the probability for the CT reaction.\\
It should be emphasised that the \textit{ab initio} study of the low-lying electronic states of the RbOH$^{-}$ system present particular challenges due to the presence of a large correlation space, and the difficulty to correctly describe the electron charge transfer, the crossing with the neutral curves and the energies at dissociation.
\FloatBarrier

\section{\label{sec:con}Conclusion}

Based on the results presented in this work, we can conclude that the rate constant of the associative detachment reaction (AD) Rb($^{2}S$)+OH$^{-}$($^{1}\Sigma^{+}$) $\rightarrow$ RbOH($^{1}\Sigma^{+}$)+e$^{-}$ strongly depends on the crossing point between the neutral and anion potential energy surfaces. This crossing point lies in the repulsive region of the PES, which is difficult to obtain accurately by quantum chemistry method. In particular, the choice of the effective core potential for the Rb atom has a drastic effect on the position of the crossing point and therefore, on the rate constant. Our results using the MWB ECP show a lack of accuracy that has already be pointed out \cite{Weigend2010} and seems to arise from an error in the ECP. We therefore recommend the use of the more recent MDF ECP when dealing with compounds containing Rb. Using the crossing point obtained with the MDF ECP along with a Langevin-based model, we found a rate constant in agreement (within the experimental uncertainty) with the experimental results of Deiglmayr \textit{et al.} \cite{Deiglmayr2012}. In addition, we found that the rate constant only decreases slightly with the temperature, which means that the reaction would also take place in the cold regime. The experimental and calculated rate constant for the AD reaction is almost 10 times smaller than the Langevin rate (4.3$\times 10^{-9}$) which implies that 10\% of the collisions are reactive. Concerning the implication for sympathetic cooling of OH$^{-}$ and the ongoing Heidelberg experiment, the AD reaction should, in principle, not prevent the feasibility of sympathetic cooling but only lead to a loss of OH$^{-}$ \cite{Deiglmayr2012}. Side effects such as collisions with vibrationally hot RbOH, product of the AD reaction, may lead to some heating processes. However this seems unlikely since the typical collision rate in such environment is on the order of tens of Hz and the time to escape the trap is in $\mu$s.\\
Collision between electronically excited Rb($^2P$) and OH$^{-}$($^1\Sigma^{+}$) are also likely to occur. The presence of conical intersections and avoided crossings in the entrance channels, as shown by our calculations, suggests that exothermic charge transfer (CT) Rb($^2P$)+OH$^{-}$($^1\Sigma^{+}$) $\rightarrow$ Rb$^{-}$($^{1}S$)+OH($^{2}\Pi$) could occur through non-adiabatic couplings. The AD reaction Rb($^2P$)+OH$^{-}$($^1\Sigma^{+}$) $\rightarrow$ RbOH($^{1}\Sigma^{+}$)+e$^{-}$ is also energetically accessible and can occur via the crossing between the excited entrance channel and the neutral PES, which delimits the autodetachment region. Our \textit{ab initio} results show that the AD reaction should prevail over the CT reaction as the system should undergo autodetachment before it can exit via the CT channel. However, a full quantum calculation would be needed in order to obtain the correct branching ratio between the CT and AD reactions. In the context of the Heidelberg experiment, the presence of Rb$^{-}$, product of the CT reaction, may be detected by the time-of-flight spectrometer if it stays trapped long enough in the rf-trap. In addition, a loss of OH$^{-}$ should be observed from the excited channels since both the AD and CT reactions induce a discharge. One would also expects this loss to be larger than the ground state since the polarizability of excited Rb($^{2}P$)(563 a.u \cite{Chen2015}) is larger than for the ground state Rb($^{2}S$)(318.6 a.u \cite{Derevianko1999}). The Langevin rate for the collision between excited Rb($^2P$) and OH$^{-}$($^1\Sigma^{+}$) becomes 6.9$\times 10^{-9}$. The observed rate constant is likely to be even larger since the induced quadrupole term have proven to be important for Rb($^{2}P$) \cite{Hall2012}. This may be seen by increasing the number of Rb atoms in their excited states, i.e increasing the intensity of the laser used in the MOT and comparing the results with the loss induced by collision with the ground state Rb($^2S$).\\ 
A comparison with other alkali atoms is currently under investigation. We will also study the effect of the spin-orbit on the low lying states of RbOH$^{-}$ and perform full quantum calculations on the dynamics. The presence of a charge transfer reaction involving anions in a cold environment offers exciting experimental opportunities and we hope that our results may help to interpret some future experimental results.  
\section*{Acknowledgments}
M.Kas is grateful to the HAItrap group of the Heidelberg University and in particular B. Höltkemeier, H. Lopez and Prof. M. Weidemüller, for their hospitality during his stays in Heidelberg and for fruitful discussions. He also wishes to thank Prof. F. A. Gianturco for helpful comments. The Fonds National de la Recherche Scientifique de Belgique (FRS-FNRS) is greatly acknowledged for financial support (FRIA grant and IISN 4.4504.10 project). We would also like to thank the ULB/VUB computing center and the CECI team for computational support.  
        
\bibliographystyle{apsrev4-1}

\begin{thebibliography}{58}%
\makeatletter
\providecommand \@ifxundefined [1]{%
 \@ifx{#1\undefined}
}%
\providecommand \@ifnum [1]{%
 \ifnum #1\expandafter \@firstoftwo
 \else \expandafter \@secondoftwo
 \fi
}%
\providecommand \@ifx [1]{%
 \ifx #1\expandafter \@firstoftwo
 \else \expandafter \@secondoftwo
 \fi
}%
\providecommand \natexlab [1]{#1}%
\providecommand \enquote  [1]{``#1''}%
\providecommand \bibnamefont  [1]{#1}%
\providecommand \bibfnamefont [1]{#1}%
\providecommand \citenamefont [1]{#1}%
\providecommand \href@noop [0]{\@secondoftwo}%
\providecommand \href [0]{\begingroup \@sanitize@url \@href}%
\providecommand \@href[1]{\@@startlink{#1}\@@href}%
\providecommand \@@href[1]{\endgroup#1\@@endlink}%
\providecommand \@sanitize@url [0]{\catcode `\\12\catcode `\$12\catcode
  `\&12\catcode `\#12\catcode `\^12\catcode `\_12\catcode `\%12\relax}%
\providecommand \@@startlink[1]{}%
\providecommand \@@endlink[0]{}%
\providecommand \url  [0]{\begingroup\@sanitize@url \@url }%
\providecommand \@url [1]{\endgroup\@href {#1}{\urlprefix }}%
\providecommand \urlprefix  [0]{URL }%
\providecommand \Eprint [0]{\href }%
\providecommand \doibase [0]{http://dx.doi.org/}%
\providecommand \selectlanguage [0]{\@gobble}%
\providecommand \bibinfo  [0]{\@secondoftwo}%
\providecommand \bibfield  [0]{\@secondoftwo}%
\providecommand \translation [1]{[#1]}%
\providecommand \BibitemOpen [0]{}%
\providecommand \bibitemStop [0]{}%
\providecommand \bibitemNoStop [0]{.\EOS\space}%
\providecommand \EOS [0]{\spacefactor3000\relax}%
\providecommand \BibitemShut  [1]{\csname bibitem#1\endcsname}%
\let\auto@bib@innerbib\@empty
\bibitem [{\citenamefont {Carr}\ \emph {et~al.}(2009)\citenamefont {Carr},
  \citenamefont {DeMille}, \citenamefont {Krems},\ and\ \citenamefont
  {Ye}}]{Carr2009}%
  \BibitemOpen
  \bibfield  {author} {\bibinfo {author} {\bibfnamefont {L.~D.}\ \bibnamefont
  {Carr}}, \bibinfo {author} {\bibfnamefont {D.}~\bibnamefont {DeMille}},
  \bibinfo {author} {\bibfnamefont {R.~V.}\ \bibnamefont {Krems}}, \ and\
  \bibinfo {author} {\bibfnamefont {J.}~\bibnamefont {Ye}},\ }\href {\doibase
  10.1088/1367-2630/11/5/055049} {\bibfield  {journal} {\bibinfo  {journal}
  {New J. Phys.}\ }\textbf {\bibinfo {volume} {11}},\ \bibinfo {pages} {055049}
  (\bibinfo {year} {2009})}\BibitemShut {NoStop}%
\bibitem [{\citenamefont {Dulieu}\ \emph {et~al.}(2011)\citenamefont {Dulieu},
  \citenamefont {Krems}, \citenamefont {Weidem{\"{u}}ller},\ and\ \citenamefont
  {Willitsch}}]{Dulieu2011}%
  \BibitemOpen
  \bibfield  {author} {\bibinfo {author} {\bibfnamefont {O.}~\bibnamefont
  {Dulieu}}, \bibinfo {author} {\bibfnamefont {R.}~\bibnamefont {Krems}},
  \bibinfo {author} {\bibfnamefont {M.}~\bibnamefont {Weidem{\"{u}}ller}}, \
  and\ \bibinfo {author} {\bibfnamefont {S.}~\bibnamefont {Willitsch}},\ }\href
  {\doibase 10.1039/c1cp90157e} {\bibfield  {journal} {\bibinfo  {journal}
  {Phys. Chem. Chem. Phys.}\ }\textbf {\bibinfo {volume} {13}},\ \bibinfo
  {pages} {18703} (\bibinfo {year} {2011})}\BibitemShut {NoStop}%
\bibitem [{\citenamefont {Krems}\ \emph {et~al.}(2010)\citenamefont {Krems},
  \citenamefont {Friedrich},\ and\ \citenamefont {Stwalley}}]{Krems2010}%
  \BibitemOpen
  \bibfield  {author} {\bibinfo {author} {\bibfnamefont {R.}~\bibnamefont
  {Krems}}, \bibinfo {author} {\bibfnamefont {B.}~\bibnamefont {Friedrich}}, \
  and\ \bibinfo {author} {\bibfnamefont {W.}~\bibnamefont {Stwalley}},\ }\href
  {http://books.google.com/books?hl=en{\&}lr={\&}id=uBqr3eOZSZYC{\&}oi=fnd{\&}pg=PR9{\&}dq=Cold+molecules,+Theory,+Experiments,+Applications{\&}ots=3zMsfkDkaL{\&}sig=47h540y4WvhOfWLHidJ9e3mC4Wo}
  {\emph {\bibinfo {title} {{Cold molecules: theory, experiment,
  applications}}}}\ (\bibinfo  {publisher} {CRC Press},\ \bibinfo {year}
  {2010})\BibitemShut {NoStop}%
\bibitem [{\citenamefont {Loh}\ \emph {et~al.}(2013)\citenamefont {Loh},
  \citenamefont {Cossel}, \citenamefont {Grau}, \citenamefont {Ni},
  \citenamefont {Meyer}, \citenamefont {Bohn}, \citenamefont {Ye},\ and\
  \citenamefont {Cornell}}]{Loh2013}%
  \BibitemOpen
  \bibfield  {author} {\bibinfo {author} {\bibfnamefont {H.}~\bibnamefont
  {Loh}}, \bibinfo {author} {\bibfnamefont {K.~C.}\ \bibnamefont {Cossel}},
  \bibinfo {author} {\bibfnamefont {M.~C.}\ \bibnamefont {Grau}}, \bibinfo
  {author} {\bibfnamefont {K.-K.}\ \bibnamefont {Ni}}, \bibinfo {author}
  {\bibfnamefont {E.~R.}\ \bibnamefont {Meyer}}, \bibinfo {author}
  {\bibfnamefont {J.~L.}\ \bibnamefont {Bohn}}, \bibinfo {author}
  {\bibfnamefont {J.}~\bibnamefont {Ye}}, \ and\ \bibinfo {author}
  {\bibfnamefont {E.~A.}\ \bibnamefont {Cornell}},\ }\href {\doibase
  10.1126/science.1243683} {\bibfield  {journal} {\bibinfo  {journal} {Science
  (80-. ).}\ }\textbf {\bibinfo {volume} {342}},\ \bibinfo {pages} {1220}
  (\bibinfo {year} {2013})}\BibitemShut {NoStop}%
\bibitem [{\citenamefont {Qu{\'{e}}m{\'{e}}ner}\ and\ \citenamefont
  {Julienne}(2012)}]{Quemener2012}%
  \BibitemOpen
  \bibfield  {author} {\bibinfo {author} {\bibfnamefont {G.}~\bibnamefont
  {Qu{\'{e}}m{\'{e}}ner}}\ and\ \bibinfo {author} {\bibfnamefont {P.~S.}\
  \bibnamefont {Julienne}},\ }\href {\doibase 10.1021/cr300092g} {\bibfield
  {journal} {\bibinfo  {journal} {Chem. Rev.}\ }\textbf {\bibinfo {volume}
  {112}},\ \bibinfo {pages} {4949} (\bibinfo {year} {2012})}\BibitemShut
  {NoStop}%
\bibitem [{\citenamefont {Wester}(2009)}]{Wester2009a}%
  \BibitemOpen
  \bibfield  {author} {\bibinfo {author} {\bibfnamefont {R.}~\bibnamefont
  {Wester}},\ }\href {\doibase 10.1088/0953-4075/42/15/154001} {\bibfield
  {journal} {\bibinfo  {journal} {J. Phys. B At. Mol. Opt. Phys.}\ }\textbf
  {\bibinfo {volume} {42}},\ \bibinfo {pages} {154001} (\bibinfo {year}
  {2009})}\BibitemShut {NoStop}%
\bibitem [{\citenamefont {Pearson}\ \emph {et~al.}(1995)\citenamefont
  {Pearson}, \citenamefont {Oesterling}, \citenamefont {Herbst},\ and\
  \citenamefont {{De Lucia}}}]{Pearson1995}%
  \BibitemOpen
  \bibfield  {author} {\bibinfo {author} {\bibfnamefont {J.~C.}\ \bibnamefont
  {Pearson}}, \bibinfo {author} {\bibfnamefont {L.~C.}\ \bibnamefont
  {Oesterling}}, \bibinfo {author} {\bibfnamefont {E.}~\bibnamefont {Herbst}},
  \ and\ \bibinfo {author} {\bibfnamefont {F.~C.}\ \bibnamefont {{De Lucia}}},\
  }\href {\doibase 10.1103/PhysRevLett.75.2940} {\bibfield  {journal} {\bibinfo
   {journal} {Phys. Rev. Lett.}\ }\textbf {\bibinfo {volume} {75}},\ \bibinfo
  {pages} {2940} (\bibinfo {year} {1995})}\BibitemShut {NoStop}%
\bibitem [{\citenamefont {Hudson}(2009)}]{Hudson2009}%
  \BibitemOpen
  \bibfield  {author} {\bibinfo {author} {\bibfnamefont {E.~R.}\ \bibnamefont
  {Hudson}},\ }\href {\doibase 10.1103/PhysRevA.79.032716} {\bibfield
  {journal} {\bibinfo  {journal} {Phys. Rev. A}\ }\textbf {\bibinfo {volume}
  {79}},\ \bibinfo {pages} {032716} (\bibinfo {year} {2009})}\BibitemShut
  {NoStop}%
\bibitem [{\citenamefont {Byrd}\ \emph {et~al.}(2013)\citenamefont {Byrd},
  \citenamefont {Michels}, \citenamefont {Montgomery},\ and\ \citenamefont
  {C{\^{o}}t{\'{e}}}}]{Byrd2013}%
  \BibitemOpen
  \bibfield  {author} {\bibinfo {author} {\bibfnamefont {J.~N.}\ \bibnamefont
  {Byrd}}, \bibinfo {author} {\bibfnamefont {H.~H.}\ \bibnamefont {Michels}},
  \bibinfo {author} {\bibfnamefont {J.~A.}\ \bibnamefont {Montgomery}}, \ and\
  \bibinfo {author} {\bibfnamefont {R.}~\bibnamefont {C{\^{o}}t{\'{e}}}},\
  }\href {\doibase 10.1103/PhysRevA.88.032710} {\bibfield  {journal} {\bibinfo
  {journal} {Phys. Rev. A}\ }\textbf {\bibinfo {volume} {88}},\ \bibinfo
  {pages} {032710} (\bibinfo {year} {2013})}\BibitemShut {NoStop}%
\bibitem [{\citenamefont {Tacconi}\ and\ \citenamefont
  {Gianturco}(2009)}]{Tacconi2009a}%
  \BibitemOpen
  \bibfield  {author} {\bibinfo {author} {\bibfnamefont {M.}~\bibnamefont
  {Tacconi}}\ and\ \bibinfo {author} {\bibfnamefont {F.~A.}\ \bibnamefont
  {Gianturco}},\ }\href {\doibase 10.1063/1.3192101} {\bibfield  {journal}
  {\bibinfo  {journal} {J. Chem. Phys.}\ }\textbf {\bibinfo {volume} {131}},\
  \bibinfo {pages} {094301} (\bibinfo {year} {2009})}\BibitemShut {NoStop}%
\bibitem [{\citenamefont {Gonz{\'{a}}lez-S{\'{a}}nchez}\ \emph
  {et~al.}(2015{\natexlab{a}})\citenamefont {Gonz{\'{a}}lez-S{\'{a}}nchez},
  \citenamefont {Carelli}, \citenamefont {Gianturco},\ and\ \citenamefont
  {Wester}}]{Gonzalez-Sanchez2015}%
  \BibitemOpen
  \bibfield  {author} {\bibinfo {author} {\bibfnamefont {L.}~\bibnamefont
  {Gonz{\'{a}}lez-S{\'{a}}nchez}}, \bibinfo {author} {\bibfnamefont
  {F.}~\bibnamefont {Carelli}}, \bibinfo {author} {\bibfnamefont
  {F.}~\bibnamefont {Gianturco}}, \ and\ \bibinfo {author} {\bibfnamefont
  {R.}~\bibnamefont {Wester}},\ }\href {\doibase
  10.1016/j.chemphys.2015.05.027} {\bibfield  {journal} {\bibinfo  {journal}
  {Chem. Phys.}\ }\textbf {\bibinfo {volume} {462}},\ \bibinfo {pages} {111}
  (\bibinfo {year} {2015}{\natexlab{a}})}\BibitemShut {NoStop}%
\bibitem [{\citenamefont {Gonzalez-Sanchez}\ \emph {et~al.}(2008)\citenamefont
  {Gonzalez-Sanchez}, \citenamefont {Tacconi}, \citenamefont {Bodo},\ and\
  \citenamefont {Gianturco}}]{Gonzalez-Sanchez2008}%
  \BibitemOpen
  \bibfield  {author} {\bibinfo {author} {\bibfnamefont {L.}~\bibnamefont
  {Gonzalez-Sanchez}}, \bibinfo {author} {\bibfnamefont {M.}~\bibnamefont
  {Tacconi}}, \bibinfo {author} {\bibfnamefont {E.}~\bibnamefont {Bodo}}, \
  and\ \bibinfo {author} {\bibfnamefont {F.~A.}\ \bibnamefont {Gianturco}},\
  }\href {\doibase 10.1140/epjd/e2008-00148-5} {\bibfield  {journal} {\bibinfo
  {journal} {Eur. Phys. J. D}\ }\textbf {\bibinfo {volume} {49}},\ \bibinfo
  {pages} {85} (\bibinfo {year} {2008})}\BibitemShut {NoStop}%
\bibitem [{\citenamefont {Gonz{\'{a}}lez-S{\'{a}}nchez}\ \emph
  {et~al.}(2015{\natexlab{b}})\citenamefont {Gonz{\'{a}}lez-S{\'{a}}nchez},
  \citenamefont {Gianturco}, \citenamefont {Carelli},\ and\ \citenamefont
  {Wester}}]{Gonzalez-Sanchez2015a}%
  \BibitemOpen
  \bibfield  {author} {\bibinfo {author} {\bibfnamefont {L.}~\bibnamefont
  {Gonz{\'{a}}lez-S{\'{a}}nchez}}, \bibinfo {author} {\bibfnamefont {F.~A.}\
  \bibnamefont {Gianturco}}, \bibinfo {author} {\bibfnamefont {F.}~\bibnamefont
  {Carelli}}, \ and\ \bibinfo {author} {\bibfnamefont {R.}~\bibnamefont
  {Wester}},\ }\href {\doibase 10.1088/1367-2630/17/12/123003} {\bibfield
  {journal} {\bibinfo  {journal} {New J. Phys.}\ }\textbf {\bibinfo {volume}
  {17}},\ \bibinfo {pages} {123003} (\bibinfo {year}
  {2015}{\natexlab{b}})}\BibitemShut {NoStop}%
\bibitem [{\citenamefont {Deiglmayr}\ \emph {et~al.}(2012)\citenamefont
  {Deiglmayr}, \citenamefont {G{\"{o}}ritz}, \citenamefont {Best},
  \citenamefont {Weidem{\"{u}}ller},\ and\ \citenamefont
  {Wester}}]{Deiglmayr2012}%
  \BibitemOpen
  \bibfield  {author} {\bibinfo {author} {\bibfnamefont {J.}~\bibnamefont
  {Deiglmayr}}, \bibinfo {author} {\bibfnamefont {A.}~\bibnamefont
  {G{\"{o}}ritz}}, \bibinfo {author} {\bibfnamefont {T.}~\bibnamefont {Best}},
  \bibinfo {author} {\bibfnamefont {M.}~\bibnamefont {Weidem{\"{u}}ller}}, \
  and\ \bibinfo {author} {\bibfnamefont {R.}~\bibnamefont {Wester}},\ }\href
  {\doibase 10.1103/PhysRevA.86.043438} {\bibfield  {journal} {\bibinfo
  {journal} {Phys. Rev. A}\ }\textbf {\bibinfo {volume} {86}},\ \bibinfo
  {pages} {043438} (\bibinfo {year} {2012})}\BibitemShut {NoStop}%
\bibitem [{\citenamefont {Otto}\ \emph {et~al.}(2008)\citenamefont {Otto},
  \citenamefont {Mikosch}, \citenamefont {Trippel}, \citenamefont
  {Weidem{\"{u}}ller},\ and\ \citenamefont {Wester}}]{Otto2008}%
  \BibitemOpen
  \bibfield  {author} {\bibinfo {author} {\bibfnamefont {R.}~\bibnamefont
  {Otto}}, \bibinfo {author} {\bibfnamefont {J.}~\bibnamefont {Mikosch}},
  \bibinfo {author} {\bibfnamefont {S.}~\bibnamefont {Trippel}}, \bibinfo
  {author} {\bibfnamefont {M.}~\bibnamefont {Weidem{\"{u}}ller}}, \ and\
  \bibinfo {author} {\bibfnamefont {R.}~\bibnamefont {Wester}},\ }\href
  {\doibase 10.1103/PhysRevLett.101.063201} {\bibfield  {journal} {\bibinfo
  {journal} {Phys. Rev. Lett.}\ }\textbf {\bibinfo {volume} {101}},\ \bibinfo
  {pages} {063201} (\bibinfo {year} {2008})}\BibitemShut {NoStop}%
\bibitem [{\citenamefont {Simons}(2008)}]{Simons2008a}%
  \BibitemOpen
  \bibfield  {author} {\bibinfo {author} {\bibfnamefont {J.}~\bibnamefont
  {Simons}},\ }\href {\doibase 10.1021/jp711490b} {\bibfield  {journal}
  {\bibinfo  {journal} {J. Phys. Chem. A}\ }\textbf {\bibinfo {volume} {112}},\
  \bibinfo {pages} {6401} (\bibinfo {year} {2008})}\BibitemShut {NoStop}%
\bibitem [{\citenamefont {Branscomb}(1966)}]{Branscomb1966}%
  \BibitemOpen
  \bibfield  {author} {\bibinfo {author} {\bibfnamefont {L.~M.}\ \bibnamefont
  {Branscomb}},\ }\href {\doibase 10.1103/PhysRev.148.11} {\bibfield  {journal}
  {\bibinfo  {journal} {Phys. Rev.}\ }\textbf {\bibinfo {volume} {148}},\
  \bibinfo {pages} {11} (\bibinfo {year} {1966})}\BibitemShut {NoStop}%
\bibitem [{\citenamefont {Hall}\ and\ \citenamefont
  {Willitsch}(2012)}]{Hall2012}%
  \BibitemOpen
  \bibfield  {author} {\bibinfo {author} {\bibfnamefont {F.~H.~J.}\
  \bibnamefont {Hall}}\ and\ \bibinfo {author} {\bibfnamefont {S.}~\bibnamefont
  {Willitsch}},\ }\href {\doibase 10.1103/PhysRevLett.109.233202} {\bibfield
  {journal} {\bibinfo  {journal} {Phys. Rev. Lett.}\ }\textbf {\bibinfo
  {volume} {109}},\ \bibinfo {pages} {233202} (\bibinfo {year}
  {2012})}\BibitemShut {NoStop}%
\bibitem [{\citenamefont {Lara}\ \emph {et~al.}(2007)\citenamefont {Lara},
  \citenamefont {Bohn}, \citenamefont {Potter}, \citenamefont {Sold{\'{a}}n},\
  and\ \citenamefont {Hutson}}]{Lara2007}%
  \BibitemOpen
  \bibfield  {author} {\bibinfo {author} {\bibfnamefont {M.}~\bibnamefont
  {Lara}}, \bibinfo {author} {\bibfnamefont {J.~L.}\ \bibnamefont {Bohn}},
  \bibinfo {author} {\bibfnamefont {D.~E.}\ \bibnamefont {Potter}}, \bibinfo
  {author} {\bibfnamefont {P.}~\bibnamefont {Sold{\'{a}}n}}, \ and\ \bibinfo
  {author} {\bibfnamefont {J.~M.}\ \bibnamefont {Hutson}},\ }\href {\doibase
  10.1103/PhysRevA.75.012704} {\bibfield  {journal} {\bibinfo  {journal} {Phys.
  Rev. A}\ }\textbf {\bibinfo {volume} {75}},\ \bibinfo {pages} {012704}
  (\bibinfo {year} {2007})}\BibitemShut {NoStop}%
\bibitem [{\citenamefont {Smith}\ \emph {et~al.}(1997)\citenamefont {Smith},
  \citenamefont {Kim},\ and\ \citenamefont {Lineberger}}]{Smith1997}%
  \BibitemOpen
  \bibfield  {author} {\bibinfo {author} {\bibfnamefont {J.~R.}\ \bibnamefont
  {Smith}}, \bibinfo {author} {\bibfnamefont {J.~B.}\ \bibnamefont {Kim}}, \
  and\ \bibinfo {author} {\bibfnamefont {W.~C.}\ \bibnamefont {Lineberger}},\
  }\href {\doibase 10.1103/PhysRevA.55.2036} {\bibfield  {journal} {\bibinfo
  {journal} {Phys. Rev. A}\ }\textbf {\bibinfo {volume} {55}},\ \bibinfo
  {pages} {2036} (\bibinfo {year} {1997})}\BibitemShut {NoStop}%
\bibitem [{\citenamefont {Frey}\ \emph {et~al.}(1978)\citenamefont {Frey},
  \citenamefont {Breyer},\ and\ \citenamefont {Holop}}]{Frey1978}%
  \BibitemOpen
  \bibfield  {author} {\bibinfo {author} {\bibfnamefont {P.}~\bibnamefont
  {Frey}}, \bibinfo {author} {\bibfnamefont {F.}~\bibnamefont {Breyer}}, \ and\
  \bibinfo {author} {\bibfnamefont {H.}~\bibnamefont {Holop}},\ }\href
  {\doibase 10.1088/0022-3700/11/19/005} {\bibfield  {journal} {\bibinfo
  {journal} {J. Phys. B At. Mol. Phys.}\ }\textbf {\bibinfo {volume} {11}},\
  \bibinfo {pages} {L589} (\bibinfo {year} {1978})}\BibitemShut {NoStop}%
\bibitem [{\citenamefont {Sansonetti}(2006)}]{Sansonetti2006}%
  \BibitemOpen
  \bibfield  {author} {\bibinfo {author} {\bibfnamefont {J.~E.}\ \bibnamefont
  {Sansonetti}},\ }\href {\doibase 10.1063/1.2035727} {\bibfield  {journal}
  {\bibinfo  {journal} {J. Phys. Chem. Ref. Data}\ }\textbf {\bibinfo {volume}
  {35}},\ \bibinfo {pages} {301} (\bibinfo {year} {2006})}\BibitemShut
  {NoStop}%
\bibitem [{\citenamefont {Werner}\ \emph {et~al.}(2012)\citenamefont {Werner},
  \citenamefont {Knowles}, \citenamefont {Knizia}, \citenamefont {Manby},
  \citenamefont {Sch{\"{u}}tz}, \citenamefont {Celani}, \citenamefont {Korona},
  \citenamefont {Lindh}, \citenamefont {Mitrushenkov}, \citenamefont {Rauhut},
  \citenamefont {Shamasundar}, \citenamefont {Adler}, \citenamefont {Amos},
  \citenamefont {Bernhardsson}, \citenamefont {Berning}, \citenamefont
  {Cooper}, \citenamefont {Deegan}, \citenamefont {Dobbyn}, \citenamefont
  {Eckert}, \citenamefont {Goll}, \citenamefont {Hampel}, \citenamefont
  {Hesselmann}, \citenamefont {Hetzer}, \citenamefont {Hrenar}, \citenamefont
  {Jansen}, \citenamefont {K{\"{o}}ppl}, \citenamefont {Liu}, \citenamefont
  {Lloyd}, \citenamefont {Mata}, \citenamefont {May}, \citenamefont
  {McNicholas}, \citenamefont {Meyer}, \citenamefont {Mura}, \citenamefont
  {Nicklass}, \citenamefont {O'Neill}, \citenamefont {Palmieri}, \citenamefont
  {Peng}, \citenamefont {Pfl{\"{u}}ger}, \citenamefont {Pitzer}, \citenamefont
  {Reiher}, \citenamefont {Shiozaki}, \citenamefont {Stoll}, \citenamefont
  {Stone}, \citenamefont {Tarroni}, \citenamefont {Thorsteinsson},\ and\
  \citenamefont {Wang}}]{MOLPRO}%
  \BibitemOpen
  \bibfield  {author} {\bibinfo {author} {\bibfnamefont {H.-J.}\ \bibnamefont
  {Werner}}, \bibinfo {author} {\bibfnamefont {P.~J.}\ \bibnamefont {Knowles}},
  \bibinfo {author} {\bibfnamefont {G.}~\bibnamefont {Knizia}}, \bibinfo
  {author} {\bibfnamefont {F.~R.}\ \bibnamefont {Manby}}, \bibinfo {author}
  {\bibfnamefont {M.}~\bibnamefont {Sch{\"{u}}tz}}, \bibinfo {author}
  {\bibfnamefont {P.}~\bibnamefont {Celani}}, \bibinfo {author} {\bibfnamefont
  {T.}~\bibnamefont {Korona}}, \bibinfo {author} {\bibfnamefont
  {R.}~\bibnamefont {Lindh}}, \bibinfo {author} {\bibfnamefont
  {A.}~\bibnamefont {Mitrushenkov}}, \bibinfo {author} {\bibfnamefont
  {G.}~\bibnamefont {Rauhut}}, \bibinfo {author} {\bibfnamefont {K.~R.}\
  \bibnamefont {Shamasundar}}, \bibinfo {author} {\bibfnamefont {T.~B.}\
  \bibnamefont {Adler}}, \bibinfo {author} {\bibfnamefont {R.~D.}\ \bibnamefont
  {Amos}}, \bibinfo {author} {\bibfnamefont {A.}~\bibnamefont {Bernhardsson}},
  \bibinfo {author} {\bibfnamefont {A.}~\bibnamefont {Berning}}, \bibinfo
  {author} {\bibfnamefont {D.~L.}\ \bibnamefont {Cooper}}, \bibinfo {author}
  {\bibfnamefont {M.~J.~O.}\ \bibnamefont {Deegan}}, \bibinfo {author}
  {\bibfnamefont {A.~J.}\ \bibnamefont {Dobbyn}}, \bibinfo {author}
  {\bibfnamefont {F.}~\bibnamefont {Eckert}}, \bibinfo {author} {\bibfnamefont
  {E.}~\bibnamefont {Goll}}, \bibinfo {author} {\bibfnamefont {C.}~\bibnamefont
  {Hampel}}, \bibinfo {author} {\bibfnamefont {A.}~\bibnamefont {Hesselmann}},
  \bibinfo {author} {\bibfnamefont {G.}~\bibnamefont {Hetzer}}, \bibinfo
  {author} {\bibfnamefont {T.}~\bibnamefont {Hrenar}}, \bibinfo {author}
  {\bibfnamefont {G.}~\bibnamefont {Jansen}}, \bibinfo {author} {\bibfnamefont
  {C.}~\bibnamefont {K{\"{o}}ppl}}, \bibinfo {author} {\bibfnamefont
  {Y.}~\bibnamefont {Liu}}, \bibinfo {author} {\bibfnamefont {A.~W.}\
  \bibnamefont {Lloyd}}, \bibinfo {author} {\bibfnamefont {R.~A.}\ \bibnamefont
  {Mata}}, \bibinfo {author} {\bibfnamefont {A.~J.}\ \bibnamefont {May}},
  \bibinfo {author} {\bibfnamefont {S.~J.}\ \bibnamefont {McNicholas}},
  \bibinfo {author} {\bibfnamefont {W.}~\bibnamefont {Meyer}}, \bibinfo
  {author} {\bibfnamefont {M.~E.}\ \bibnamefont {Mura}}, \bibinfo {author}
  {\bibfnamefont {A.}~\bibnamefont {Nicklass}}, \bibinfo {author}
  {\bibfnamefont {D.~P.}\ \bibnamefont {O'Neill}}, \bibinfo {author}
  {\bibfnamefont {P.}~\bibnamefont {Palmieri}}, \bibinfo {author}
  {\bibfnamefont {D.}~\bibnamefont {Peng}}, \bibinfo {author} {\bibfnamefont
  {K.}~\bibnamefont {Pfl{\"{u}}ger}}, \bibinfo {author} {\bibfnamefont
  {R.}~\bibnamefont {Pitzer}}, \bibinfo {author} {\bibfnamefont
  {M.}~\bibnamefont {Reiher}}, \bibinfo {author} {\bibfnamefont
  {T.}~\bibnamefont {Shiozaki}}, \bibinfo {author} {\bibfnamefont
  {H.}~\bibnamefont {Stoll}}, \bibinfo {author} {\bibfnamefont {A.~J.}\
  \bibnamefont {Stone}}, \bibinfo {author} {\bibfnamefont {R.}~\bibnamefont
  {Tarroni}}, \bibinfo {author} {\bibfnamefont {T.}~\bibnamefont
  {Thorsteinsson}}, \ and\ \bibinfo {author} {\bibfnamefont {M.}~\bibnamefont
  {Wang}},\ }\href@noop {} {\enquote {\bibinfo {title} {{MOLPRO, version
  2012.1, a package of ab initio programs}},}\ } (\bibinfo {year}
  {2012})\BibitemShut {NoStop}%
\bibitem [{\citenamefont {Hampel}\ \emph {et~al.}(1992)\citenamefont {Hampel},
  \citenamefont {Peterson},\ and\ \citenamefont {Werner}}]{Hampel1992}%
  \BibitemOpen
  \bibfield  {author} {\bibinfo {author} {\bibfnamefont {C.}~\bibnamefont
  {Hampel}}, \bibinfo {author} {\bibfnamefont {K.~a.}\ \bibnamefont
  {Peterson}}, \ and\ \bibinfo {author} {\bibfnamefont {H.-J.}\ \bibnamefont
  {Werner}},\ }\href {\doibase 10.1016/0009-2614(92)86093-W} {\bibfield
  {journal} {\bibinfo  {journal} {Chem. Phys. Lett.}\ }\textbf {\bibinfo
  {volume} {190}},\ \bibinfo {pages} {1} (\bibinfo {year} {1992})}\BibitemShut
  {NoStop}%
\bibitem [{\citenamefont {Knowles}\ \emph {et~al.}(1993)\citenamefont
  {Knowles}, \citenamefont {Hampel},\ and\ \citenamefont
  {Werner}}]{Knowles1993}%
  \BibitemOpen
  \bibfield  {author} {\bibinfo {author} {\bibfnamefont {P.~J.}\ \bibnamefont
  {Knowles}}, \bibinfo {author} {\bibfnamefont {C.}~\bibnamefont {Hampel}}, \
  and\ \bibinfo {author} {\bibfnamefont {H.-J.}\ \bibnamefont {Werner}},\
  }\href {\doibase 10.1063/1.465990} {\bibfield  {journal} {\bibinfo  {journal}
  {J. Chem. Phys.}\ }\textbf {\bibinfo {volume} {99}},\ \bibinfo {pages} {5219}
  (\bibinfo {year} {1993})}\BibitemShut {NoStop}%
\bibitem [{\citenamefont {Knowles}\ \emph {et~al.}(2000)\citenamefont
  {Knowles}, \citenamefont {Hampel},\ and\ \citenamefont
  {Werner}}]{Knowles2000}%
  \BibitemOpen
  \bibfield  {author} {\bibinfo {author} {\bibfnamefont {P.~J.}\ \bibnamefont
  {Knowles}}, \bibinfo {author} {\bibfnamefont {C.}~\bibnamefont {Hampel}}, \
  and\ \bibinfo {author} {\bibfnamefont {H.-J.}\ \bibnamefont {Werner}},\
  }\href {\doibase 10.1063/1.480886} {\bibfield  {journal} {\bibinfo  {journal}
  {J. Chem. Phys.}\ }\textbf {\bibinfo {volume} {112}},\ \bibinfo {pages}
  {3106} (\bibinfo {year} {2000})}\BibitemShut {NoStop}%
\bibitem [{\citenamefont {Dunning}(1989)}]{ThomH.Dunning1988}%
  \BibitemOpen
  \bibfield  {author} {\bibinfo {author} {\bibfnamefont {T.~H.}\ \bibnamefont
  {Dunning}},\ }\href {\doibase 10.1063/1.456153} {\bibfield  {journal}
  {\bibinfo  {journal} {J. Chem. Phys.}\ }\textbf {\bibinfo {volume} {90}},\
  \bibinfo {pages} {1007} (\bibinfo {year} {1989})}\BibitemShut {NoStop}%
\bibitem [{\citenamefont {Leininger}\ \emph {et~al.}(1996)\citenamefont
  {Leininger}, \citenamefont {Nicklass}, \citenamefont {K{\"{u}}chle},
  \citenamefont {Stoll}, \citenamefont {Dolg},\ and\ \citenamefont
  {Bergner}}]{Leininger1996}%
  \BibitemOpen
  \bibfield  {author} {\bibinfo {author} {\bibfnamefont {T.}~\bibnamefont
  {Leininger}}, \bibinfo {author} {\bibfnamefont {A.}~\bibnamefont {Nicklass}},
  \bibinfo {author} {\bibfnamefont {W.}~\bibnamefont {K{\"{u}}chle}}, \bibinfo
  {author} {\bibfnamefont {H.}~\bibnamefont {Stoll}}, \bibinfo {author}
  {\bibfnamefont {M.}~\bibnamefont {Dolg}}, \ and\ \bibinfo {author}
  {\bibfnamefont {A.}~\bibnamefont {Bergner}},\ }\href {\doibase
  10.1016/0009-2614(96)00382-X} {\bibfield  {journal} {\bibinfo  {journal}
  {Chem. Phys. Lett.}\ }\textbf {\bibinfo {volume} {255}},\ \bibinfo {pages}
  {274} (\bibinfo {year} {1996})}\BibitemShut {NoStop}%
\bibitem [{\citenamefont {Weigend}\ \emph {et~al.}(2003)\citenamefont
  {Weigend}, \citenamefont {Furche},\ and\ \citenamefont
  {Ahlrichs}}]{Weigend2003}%
  \BibitemOpen
  \bibfield  {author} {\bibinfo {author} {\bibfnamefont {F.}~\bibnamefont
  {Weigend}}, \bibinfo {author} {\bibfnamefont {F.}~\bibnamefont {Furche}}, \
  and\ \bibinfo {author} {\bibfnamefont {R.}~\bibnamefont {Ahlrichs}},\ }\href
  {\doibase 10.1063/1.1627293} {\bibfield  {journal} {\bibinfo  {journal} {J.
  Chem. Phys.}\ }\textbf {\bibinfo {volume} {119}},\ \bibinfo {pages} {12753}
  (\bibinfo {year} {2003})}\BibitemShut {NoStop}%
\bibitem [{\citenamefont {Weigend}\ and\ \citenamefont
  {Ahlrichs}(2005)}]{Weigend2005}%
  \BibitemOpen
  \bibfield  {author} {\bibinfo {author} {\bibfnamefont {F.}~\bibnamefont
  {Weigend}}\ and\ \bibinfo {author} {\bibfnamefont {R.}~\bibnamefont
  {Ahlrichs}},\ }\href {\doibase 10.1039/b508541a} {\bibfield  {journal}
  {\bibinfo  {journal} {Phys. Chem. Chem. Phys.}\ }\textbf {\bibinfo {volume}
  {7}},\ \bibinfo {pages} {3297} (\bibinfo {year} {2005})}\BibitemShut
  {NoStop}%
\bibitem [{\citenamefont {Lim}\ \emph {et~al.}(2005)\citenamefont {Lim},
  \citenamefont {Schwerdtfeger}, \citenamefont {Metz},\ and\ \citenamefont
  {Stoll}}]{Lim2005}%
  \BibitemOpen
  \bibfield  {author} {\bibinfo {author} {\bibfnamefont {I.~S.}\ \bibnamefont
  {Lim}}, \bibinfo {author} {\bibfnamefont {P.}~\bibnamefont {Schwerdtfeger}},
  \bibinfo {author} {\bibfnamefont {B.}~\bibnamefont {Metz}}, \ and\ \bibinfo
  {author} {\bibfnamefont {H.}~\bibnamefont {Stoll}},\ }\href {\doibase
  10.1063/1.1856451} {\bibfield  {journal} {\bibinfo  {journal} {J. Chem.
  Phys.}\ }\textbf {\bibinfo {volume} {122}},\ \bibinfo {pages} {104103}
  (\bibinfo {year} {2005})}\BibitemShut {NoStop}%
\bibitem [{\citenamefont {Weigend}\ and\ \citenamefont
  {Baldes}(2010)}]{Weigend2010}%
  \BibitemOpen
  \bibfield  {author} {\bibinfo {author} {\bibfnamefont {F.}~\bibnamefont
  {Weigend}}\ and\ \bibinfo {author} {\bibfnamefont {A.}~\bibnamefont
  {Baldes}},\ }\href {\doibase 10.1063/1.3495681} {\bibfield  {journal}
  {\bibinfo  {journal} {J. Chem. Phys.}\ }\textbf {\bibinfo {volume} {133}},\
  \bibinfo {pages} {174102} (\bibinfo {year} {2010})}\BibitemShut {NoStop}%
\bibitem [{\citenamefont {Boys}\ and\ \citenamefont
  {Bernardi}(1970)}]{Boys1970}%
  \BibitemOpen
  \bibfield  {author} {\bibinfo {author} {\bibfnamefont {S.}~\bibnamefont
  {Boys}}\ and\ \bibinfo {author} {\bibfnamefont {F.}~\bibnamefont
  {Bernardi}},\ }\href {\doibase 10.1080/00268977000101561} {\bibfield
  {journal} {\bibinfo  {journal} {Mol. Phys.}\ }\textbf {\bibinfo {volume}
  {19}},\ \bibinfo {pages} {553} (\bibinfo {year} {1970})}\BibitemShut
  {NoStop}%
\bibitem [{\citenamefont {Knowles}\ \emph {et~al.}(1991)\citenamefont
  {Knowles}, \citenamefont {Andrews}, \citenamefont {Amos}, \citenamefont
  {Handy},\ and\ \citenamefont {Pople}}]{KAAHP91}%
  \BibitemOpen
  \bibfield  {author} {\bibinfo {author} {\bibfnamefont {P.~J.}\ \bibnamefont
  {Knowles}}, \bibinfo {author} {\bibfnamefont {J.~S.}\ \bibnamefont
  {Andrews}}, \bibinfo {author} {\bibfnamefont {R.~D.}\ \bibnamefont {Amos}},
  \bibinfo {author} {\bibfnamefont {N.~C.}\ \bibnamefont {Handy}}, \ and\
  \bibinfo {author} {\bibfnamefont {J.~A.}\ \bibnamefont {Pople}},\ }\href
  {\doibase 10.1016/S0009-2614(91)85118-G} {\bibfield  {journal} {\bibinfo
  {journal} {Chem. Phys. Lett.}\ }\textbf {\bibinfo {volume} {186}},\ \bibinfo
  {pages} {130} (\bibinfo {year} {1991})}\BibitemShut {NoStop}%
\bibitem [{\citenamefont {Amos}\ \emph {et~al.}(1991)\citenamefont {Amos},
  \citenamefont {Andrews}, \citenamefont {Handy},\ and\ \citenamefont
  {Knowles}}]{AAHK91}%
  \BibitemOpen
  \bibfield  {author} {\bibinfo {author} {\bibfnamefont {R.~D.}\ \bibnamefont
  {Amos}}, \bibinfo {author} {\bibfnamefont {J.~S.}\ \bibnamefont {Andrews}},
  \bibinfo {author} {\bibfnamefont {N.~C.}\ \bibnamefont {Handy}}, \ and\
  \bibinfo {author} {\bibfnamefont {P.~J.}\ \bibnamefont {Knowles}},\ }\href
  {\doibase 10.1016/S0009-2614(91)85057-4} {\bibfield  {journal} {\bibinfo
  {journal} {Chem. Phys. Lett.}\ }\textbf {\bibinfo {volume} {185}},\ \bibinfo
  {pages} {256} (\bibinfo {year} {1991})}\BibitemShut {NoStop}%
\bibitem [{\citenamefont {Kat{\^{o}}}\ \emph {et~al.}(1985)\citenamefont
  {Kat{\^{o}}}, \citenamefont {Toyosaka},\ and\ \citenamefont
  {Suzuki}}]{Kato1985}%
  \BibitemOpen
  \bibfield  {author} {\bibinfo {author} {\bibfnamefont {H.}~\bibnamefont
  {Kat{\^{o}}}}, \bibinfo {author} {\bibfnamefont {Y.}~\bibnamefont
  {Toyosaka}}, \ and\ \bibinfo {author} {\bibfnamefont {T.}~\bibnamefont
  {Suzuki}},\ }\href {\doibase 10.1246/bcsj.58.562} {\bibfield  {journal}
  {\bibinfo  {journal} {Bull. Chem. Soc. Jpn.}\ }\textbf {\bibinfo {volume}
  {58}},\ \bibinfo {pages} {562} (\bibinfo {year} {1985})}\BibitemShut
  {NoStop}%
\bibitem [{\citenamefont {Rice}\ and\ \citenamefont
  {Klemperer}(1957)}]{Rice1957}%
  \BibitemOpen
  \bibfield  {author} {\bibinfo {author} {\bibfnamefont {S.~a.}\ \bibnamefont
  {Rice}}\ and\ \bibinfo {author} {\bibfnamefont {W.}~\bibnamefont
  {Klemperer}},\ }\href {\doibase 10.1063/1.1743772} {\bibfield  {journal}
  {\bibinfo  {journal} {J. Chem. Phys.}\ }\textbf {\bibinfo {volume} {27}},\
  \bibinfo {pages} {573} (\bibinfo {year} {1957})}\BibitemShut {NoStop}%
\bibitem [{\citenamefont {{Bonczyk P.A. and Hughes V.W}}(1967)}]{Hughes1967}%
  \BibitemOpen
  \bibfield  {author} {\bibinfo {author} {\bibnamefont {{Bonczyk P.A. and
  Hughes V.W}}},\ }\href {\doibase 0.1103/PhysRev.161.15} {\bibfield  {journal}
  {\bibinfo  {journal} {Rev. Phys.}\ }\textbf {\bibinfo {volume} {161}}
  (\bibinfo {year} {1967}),\ 0.1103/PhysRev.161.15}\BibitemShut {NoStop}%
\bibitem [{\citenamefont {Yamada}\ and\ \citenamefont
  {Hirota}(1999)}]{Yamada1999}%
  \BibitemOpen
  \bibfield  {author} {\bibinfo {author} {\bibfnamefont {C.}~\bibnamefont
  {Yamada}}\ and\ \bibinfo {author} {\bibfnamefont {E.}~\bibnamefont
  {Hirota}},\ }\href {\doibase 10.1063/1.477927} {\bibfield  {journal}
  {\bibinfo  {journal} {J. Chem. Phys.}\ }\textbf {\bibinfo {volume} {110}},\
  \bibinfo {pages} {2853} (\bibinfo {year} {1999})}\BibitemShut {NoStop}%
\bibitem [{\citenamefont {Simons}(1998)}]{Simons1998}%
  \BibitemOpen
  \bibfield  {author} {\bibinfo {author} {\bibfnamefont {J.}~\bibnamefont
  {Simons}},\ }\href {\doibase 10.1021/jp981663u} {\bibfield  {journal}
  {\bibinfo  {journal} {J. Phys. Chem. A}\ }\textbf {\bibinfo {volume} {102}},\
  \bibinfo {pages} {6035} (\bibinfo {year} {1998})}\BibitemShut {NoStop}%
\bibitem [{\citenamefont {Acharya}\ \emph {et~al.}(1985)\citenamefont
  {Acharya}, \citenamefont {Kendall},\ and\ \citenamefont
  {Simons}}]{Acharya1985}%
  \BibitemOpen
  \bibfield  {author} {\bibinfo {author} {\bibfnamefont {P.~K.}\ \bibnamefont
  {Acharya}}, \bibinfo {author} {\bibfnamefont {R.~a.}\ \bibnamefont
  {Kendall}}, \ and\ \bibinfo {author} {\bibfnamefont {J.}~\bibnamefont
  {Simons}},\ }\href {\doibase 10.1063/1.449100} {\bibfield  {journal}
  {\bibinfo  {journal} {J. Chem. Phys.}\ }\textbf {\bibinfo {volume} {83}},\
  \bibinfo {pages} {3888} (\bibinfo {year} {1985})}\BibitemShut {NoStop}%
\bibitem [{\citenamefont {Fedchak}\ \emph {et~al.}(1993)\citenamefont
  {Fedchak}, \citenamefont {Huels}, \citenamefont {Doverspike},\ and\
  \citenamefont {Champion}}]{Fedchak1993}%
  \BibitemOpen
  \bibfield  {author} {\bibinfo {author} {\bibfnamefont {J.~A.}\ \bibnamefont
  {Fedchak}}, \bibinfo {author} {\bibfnamefont {M.~A.}\ \bibnamefont {Huels}},
  \bibinfo {author} {\bibfnamefont {L.~D.}\ \bibnamefont {Doverspike}}, \ and\
  \bibinfo {author} {\bibfnamefont {R.~L.}\ \bibnamefont {Champion}},\ }\href
  {\doibase 10.1103/PhysRevA.47.3796} {\bibfield  {journal} {\bibinfo
  {journal} {Phys. Rev. A}\ }\textbf {\bibinfo {volume} {47}},\ \bibinfo
  {pages} {3796} (\bibinfo {year} {1993})}\BibitemShut {NoStop}%
\bibitem [{\citenamefont {{\v{C}}{\'{\i}}{\v{z}}ek}\ \emph
  {et~al.}(1999)\citenamefont {{\v{C}}{\'{\i}}{\v{z}}ek}, \citenamefont
  {Hor{\'{a}}{\v{c}}ek},\ and\ \citenamefont {Domcke}}]{Cizek1999}%
  \BibitemOpen
  \bibfield  {author} {\bibinfo {author} {\bibfnamefont {M.}~\bibnamefont
  {{\v{C}}{\'{\i}}{\v{z}}ek}}, \bibinfo {author} {\bibfnamefont
  {J.}~\bibnamefont {Hor{\'{a}}{\v{c}}ek}}, \ and\ \bibinfo {author}
  {\bibfnamefont {W.}~\bibnamefont {Domcke}},\ }\href {\doibase
  10.1103/PhysRevA.60.2873} {\bibfield  {journal} {\bibinfo  {journal} {Phys.
  Rev. A}\ }\textbf {\bibinfo {volume} {60}},\ \bibinfo {pages} {2873}
  (\bibinfo {year} {1999})}\BibitemShut {NoStop}%
\bibitem [{\citenamefont {C{\'{\i}}zek}\ \emph {et~al.}(2001)\citenamefont
  {C{\'{\i}}zek}, \citenamefont {Hor{\'{a}}cek}, \citenamefont {Thiel},\ and\
  \citenamefont {Hotop}}]{Cizek2001}%
  \BibitemOpen
  \bibfield  {author} {\bibinfo {author} {\bibfnamefont {M.}~\bibnamefont
  {C{\'{\i}}zek}}, \bibinfo {author} {\bibfnamefont {J.}~\bibnamefont
  {Hor{\'{a}}cek}}, \bibinfo {author} {\bibfnamefont {F.~A.~U.}\ \bibnamefont
  {Thiel}}, \ and\ \bibinfo {author} {\bibfnamefont {H.}~\bibnamefont
  {Hotop}},\ }\href {\doibase 10.1088/0953-4075/34/6/302} {\bibfield  {journal}
  {\bibinfo  {journal} {J. Phys. B At. Mol. Opt. Phys.}\ }\textbf {\bibinfo
  {volume} {34}},\ \bibinfo {pages} {983} (\bibinfo {year} {2001})}\BibitemShut
  {NoStop}%
\bibitem [{\citenamefont {Domcke}\ and\ \citenamefont
  {Mundel}(1985)}]{Domcke1985}%
  \BibitemOpen
  \bibfield  {author} {\bibinfo {author} {\bibfnamefont {W.}~\bibnamefont
  {Domcke}}\ and\ \bibinfo {author} {\bibfnamefont {C.}~\bibnamefont
  {Mundel}},\ }\href {\doibase 10.1088/0022-3700/18/22/017} {\bibfield
  {journal} {\bibinfo  {journal} {J. Phys. B At. Mol. Phys.}\ }\textbf
  {\bibinfo {volume} {18}},\ \bibinfo {pages} {4491} (\bibinfo {year}
  {1985})}\BibitemShut {NoStop}%
\bibitem [{\citenamefont {Schulz}\ and\ \citenamefont
  {G.J.}(1973)}]{Schultz1973}%
  \BibitemOpen
  \bibfield  {author} {\bibinfo {author} {\bibfnamefont {J.~M.}\ \bibnamefont
  {Schulz}}\ and\ \bibinfo {author} {\bibnamefont {G.J.}},\ }\href@noop {}
  {\bibfield  {journal} {\bibinfo  {journal} {Phys. Rev. A}\ }\textbf {\bibinfo
  {volume} {7}} (\bibinfo {year} {1973})}\BibitemShut {NoStop}%
\bibitem [{\citenamefont {Derevianko}\ \emph {et~al.}(1999)\citenamefont
  {Derevianko}, \citenamefont {Johnson}, \citenamefont {Safronova},\ and\
  \citenamefont {Babb}}]{Derevianko1999}%
  \BibitemOpen
  \bibfield  {author} {\bibinfo {author} {\bibfnamefont {a.}~\bibnamefont
  {Derevianko}}, \bibinfo {author} {\bibfnamefont {W.~R.}\ \bibnamefont
  {Johnson}}, \bibinfo {author} {\bibfnamefont {M.~S.}\ \bibnamefont
  {Safronova}}, \ and\ \bibinfo {author} {\bibfnamefont {J.~F.}\ \bibnamefont
  {Babb}},\ }\href {\doibase 10.1103/PhysRevLett.82.3589} {\bibfield  {journal}
  {\bibinfo  {journal} {Phys. Rev. Lett.}\ }\textbf {\bibinfo {volume} {82}},\
  \bibinfo {pages} {3589} (\bibinfo {year} {1999})},\ \Eprint
  {http://arxiv.org/abs/9812028} {9812028 [physics]} \BibitemShut {NoStop}%
\bibitem [{\citenamefont {Rosenbaum}\ \emph {et~al.}(1986)\citenamefont
  {Rosenbaum}, \citenamefont {Owrutsky}, \citenamefont {Tack},\ and\
  \citenamefont {Saykally}}]{Rosenbaum1986}%
  \BibitemOpen
  \bibfield  {author} {\bibinfo {author} {\bibfnamefont {N.~H.}\ \bibnamefont
  {Rosenbaum}}, \bibinfo {author} {\bibfnamefont {J.~C.}\ \bibnamefont
  {Owrutsky}}, \bibinfo {author} {\bibfnamefont {L.~M.}\ \bibnamefont {Tack}},
  \ and\ \bibinfo {author} {\bibfnamefont {R.~J.}\ \bibnamefont {Saykally}},\
  }\href {\doibase 10.1063/1.449941} {\bibfield  {journal} {\bibinfo  {journal}
  {J. Chem. Phys.}\ }\textbf {\bibinfo {volume} {84}},\ \bibinfo {pages} {5308}
  (\bibinfo {year} {1986})}\BibitemShut {NoStop}%
\bibitem [{\citenamefont {Knowles}\ and\ \citenamefont
  {Werner}(1992)}]{Knowles1992}%
  \BibitemOpen
  \bibfield  {author} {\bibinfo {author} {\bibfnamefont {P.~J.}\ \bibnamefont
  {Knowles}}\ and\ \bibinfo {author} {\bibfnamefont {H.-J.}\ \bibnamefont
  {Werner}},\ }\href {\doibase 10.1007/BF01117405} {\bibfield  {journal}
  {\bibinfo  {journal} {Theor. Chim. Acta}\ }\textbf {\bibinfo {volume} {84}},\
  \bibinfo {pages} {95} (\bibinfo {year} {1992})}\BibitemShut {NoStop}%
\bibitem [{\citenamefont {Langhoff}\ and\ \citenamefont
  {Davidson}(1974)}]{Langhoff1974}%
  \BibitemOpen
  \bibfield  {author} {\bibinfo {author} {\bibfnamefont {S.~R.}\ \bibnamefont
  {Langhoff}}\ and\ \bibinfo {author} {\bibfnamefont {E.~R.}\ \bibnamefont
  {Davidson}},\ }\href {\doibase 10.1002/qua.560080106} {\bibfield  {journal}
  {\bibinfo  {journal} {Int. J. Quantum Chem.}\ }\textbf {\bibinfo {volume}
  {8}},\ \bibinfo {pages} {61} (\bibinfo {year} {1974})}\BibitemShut {NoStop}%
\bibitem [{\citenamefont {Meissner}(1988)}]{Meissner1988}%
  \BibitemOpen
  \bibfield  {author} {\bibinfo {author} {\bibfnamefont {L.}~\bibnamefont
  {Meissner}},\ }\href {\doibase 10.1016/0009-2614(88)87431-1} {\bibfield
  {journal} {\bibinfo  {journal} {Chem. Phys. Lett.}\ }\textbf {\bibinfo
  {volume} {146}},\ \bibinfo {pages} {204} (\bibinfo {year}
  {1988})}\BibitemShut {NoStop}%
\bibitem [{\citenamefont {Werner}\ \emph {et~al.}(2008)\citenamefont {Werner},
  \citenamefont {Kállay},\ and\ \citenamefont {Gauss}}]{Werner2008}%
  \BibitemOpen
  \bibfield  {author} {\bibinfo {author} {\bibfnamefont {H.-J.}\ \bibnamefont
  {Werner}}, \bibinfo {author} {\bibfnamefont {M.}~\bibnamefont {K\'{a}llay}}, \
  and\ \bibinfo {author} {\bibfnamefont {J.}~\bibnamefont {Gauss}},\ }\href
  {\doibase 10.1063/1.2822905} {\bibfield  {journal} {\bibinfo  {journal} {J.
  Chem. Phys.}\ }\textbf {\bibinfo {volume} {128}},\ \bibinfo {pages} {034305}
  (\bibinfo {year} {2008})}\BibitemShut {NoStop}%
\bibitem [{\citenamefont {Jahn}\ and\ \citenamefont {Teller}(1937)}]{Jahn1937}%
  \BibitemOpen
  \bibfield  {author} {\bibinfo {author} {\bibfnamefont {H.~A.}\ \bibnamefont
  {Jahn}}\ and\ \bibinfo {author} {\bibfnamefont {E.}~\bibnamefont {Teller}},\
  }\href {\doibase 10.1098/rspa.1937.0142} {\bibfield  {journal} {\bibinfo
  {journal} {Proc. R. Soc. A Math. Phys. Eng. Sci.}\ }\textbf {\bibinfo
  {volume} {161}},\ \bibinfo {pages} {220} (\bibinfo {year}
  {1937})}\BibitemShut {NoStop}%
\bibitem [{\citenamefont {Pearson}(1975)}]{Pearson1975}%
  \BibitemOpen
  \bibfield  {author} {\bibinfo {author} {\bibfnamefont {R.~G.}\ \bibnamefont
  {Pearson}},\ }\href {\doibase 10.1073/pnas.72.6.2104} {\bibfield  {journal}
  {\bibinfo  {journal} {Proc. Natl. Acad. Sci.}\ }\textbf {\bibinfo {volume}
  {72}},\ \bibinfo {pages} {2104} (\bibinfo {year} {1975})}\BibitemShut
  {NoStop}%
\bibitem [{\citenamefont {Rienstra-Kiracofe}\ \emph {et~al.}(2002)\citenamefont
  {Rienstra-Kiracofe}, \citenamefont {Tschumper}, \citenamefont {Schaefer},
  \citenamefont {Nandi},\ and\ \citenamefont
  {Ellison}}]{Rienstra-Kiracofe2002}%
  \BibitemOpen
  \bibfield  {author} {\bibinfo {author} {\bibfnamefont {J.~C.}\ \bibnamefont
  {Rienstra-Kiracofe}}, \bibinfo {author} {\bibfnamefont {G.~S.}\ \bibnamefont
  {Tschumper}}, \bibinfo {author} {\bibfnamefont {H.~F.}\ \bibnamefont
  {Schaefer}}, \bibinfo {author} {\bibfnamefont {S.}~\bibnamefont {Nandi}}, \
  and\ \bibinfo {author} {\bibfnamefont {G.~B.}\ \bibnamefont {Ellison}},\
  }\href {\doibase 10.1021/cr990044u} {\bibfield  {journal} {\bibinfo
  {journal} {Chem. Rev.}\ }\textbf {\bibinfo {volume} {102}},\ \bibinfo {pages}
  {231} (\bibinfo {year} {2002})}\BibitemShut {NoStop}%
\bibitem [{\citenamefont {Feller}\ and\ \citenamefont
  {Davidson}(1989)}]{Feller1989}%
  \BibitemOpen
  \bibfield  {author} {\bibinfo {author} {\bibfnamefont {D.}~\bibnamefont
  {Feller}}\ and\ \bibinfo {author} {\bibfnamefont {E.~R.}\ \bibnamefont
  {Davidson}},\ }\href {\doibase 10.1063/1.456154} {\bibfield  {journal}
  {\bibinfo  {journal} {J. Chem. Phys.}\ }\textbf {\bibinfo {volume} {90}},\
  \bibinfo {pages} {1024} (\bibinfo {year} {1989})}\BibitemShut {NoStop}%
\bibitem [{\citenamefont {Martin}(2001)}]{Martin2001}%
  \BibitemOpen
  \bibfield  {author} {\bibinfo {author} {\bibfnamefont {J.~M.}\ \bibnamefont
  {Martin}},\ }\href {\doibase 10.1016/S1386-1425(00)00450-9} {\bibfield
  {journal} {\bibinfo  {journal} {Spectrochim. Acta Part A Mol. Biomol.
  Spectrosc.}\ }\textbf {\bibinfo {volume} {57}},\ \bibinfo {pages} {875}
  (\bibinfo {year} {2001})}\BibitemShut {NoStop}%
\bibitem [{\citenamefont {Chen}\ \emph {et~al.}(2015)\citenamefont {Chen},
  \citenamefont {Gon{\c{c}}alves},\ and\ \citenamefont {Raithel}}]{Chen2015}%
  \BibitemOpen
  \bibfield  {author} {\bibinfo {author} {\bibfnamefont {Y.-J.}\ \bibnamefont
  {Chen}}, \bibinfo {author} {\bibfnamefont {L.~F.}\ \bibnamefont
  {Gon{\c{c}}alves}}, \ and\ \bibinfo {author} {\bibfnamefont {G.}~\bibnamefont
  {Raithel}},\ }\href {\doibase 10.1103/PhysRevA.92.060501} {\bibfield
  {journal} {\bibinfo  {journal} {Phys. Rev. A}\ }\textbf {\bibinfo {volume}
  {92}},\ \bibinfo {pages} {060501} (\bibinfo {year} {2015})}\BibitemShut
  {NoStop}%
\end{thebibliography}
%
\end{document}